%% file: petropoulouMSRevi.tex
\documentclass[preprint2]{aastex}

\newcommand{\HII}{H {\scshape{ii}} }
\newcommand{\OII}{[O {\scshape{ii}}] }
\newcommand{\OIII}{[O {\scshape{iii}}] }
\newcommand{\SII}{[S {\scshape{ii}}] }
\newcommand{\NII}{[N {\scshape{ii}}] }
\newcommand{\Ha}{H$\alpha$ }

\slugcomment{Not to appear in Nonlearned J., 45.}
\shorttitle{}
\shortauthors{Petropoulou et al.}

\begin{document}

\title{Environmental Effects on the Metal Enrichment of Low Mass  Galaxies in Nearby Clusters}

\author{Petropoulou V.\altaffilmark{1},V\'ilchez J.\altaffilmark{1}, Iglesias-P\'aramo J.\altaffilmark{1,2}}
\affil{\altaffilmark{1}Instituto de Astrof\'isica de Andaluc\'ia- C.S.I.C., Glorieta de la Astronom\'ia, 18008 Granada, Spain;\\
\altaffilmark{2}Centro Astron\'omico Hispano Alem\'an, C/ Jes\'us Durb\'an Rem\'on 2-2, 04004 Almer\'ia, Spain;\\
}

\begin{abstract}

In this paper we study the chemical history of low-mass star-forming (SF) galaxies in the local Universe clusters Coma, A1367, A779, and A634. The aim of this work is to search for the imprint of the environment on the chemical evolution of these galaxies.  Galaxy chemical evolution is linked to the star formation history (SFH), as well as to the gas interchange with the environment, and low-mass galaxies are well known to be vulnerable systems to environmental processes affecting both these parameters. For our study we have used spectra from the SDSS-III DR8. We have examined the spectroscopic properties of  SF galaxies of stellar masses $10^8-10^{10} M_\sun$, located from the core to the cluster's outskirts. The gas-phase O/H and N/O chemical abundances have been derived using the latest empirical calibrations.
We have examined the mass-metallicity relation of cluster galaxies finding well defined sequences. The slope of these sequences, for galaxies in low-mass clusters and galaxies at large cluster-centric distances, follows the predictions of recent hydrodynamic models. A flattening of this slope has been observed for galaxies located in the core of the two more massive clusters of the sample, principally in Coma, suggesting that the imprint of the cluster environment on the chemical evolution of SF galaxies should be sensitive to both the galaxy mass and the host cluster mass. The HI gas content of Coma and A1367 galaxies indicate that low-mass SF galaxies, located at the core of these clusters, have been severely affected by ram-pressure stripping. The observed mass-dependent enhancement of the metal content of low-mass galaxies in dense environments seems plausible, according to hydrodynamic simulations. This enhanced metal enrichment could be produced by the combination of effects such as wind reaccretion, due to pressure cofinement by the intra-cluster medium (ICM), and the truncation of gas infall, as a result of the ram-pressure stripping. Thus, the properties of the ICM should play an important role in the chemical evolution of low-mass galaxies in clusters.

\end{abstract}


\section{INTRODUCTION}\label{INTRO}
The chemical evolution of a galaxy is linked to the SFH, as well as to the gas interchange with the environment via inflows or outflows. In this sense, galaxy metallicity,  could be an observable parameter providing information on the impact of the environment on the galaxy SFH and/or the galaxy gas content. 

SFH has been found to be related to the environment in which a galaxy inhabits: dense regions such as galaxy clusters, are mainly populated by red passive spheroids, whereas SF galaxies dominate in the field \citep{Dressler1980, Balogh1999, Poggianti1999, Treu2003, Balogh2004, Rines2005, Finn2005, Haines2007, Poggianti2008}. Though, intense discussions on the underlying cause of this observational trend are still ongoing. One interpretation has been that this trend could be the mere result of the fact that high density regions have favored the formation mostly of massive galaxies, which in turn convert their gas in stars faster than dwarf galaxies \citep{Kennicutt1998} and they have already finished forming stars earlier in the past \citep{Merlin2006}. Recent observational evidences support the idea that the stellar mass function can be associated to the environment \citep{Vulcani2011,Bolzonella2010}. 

Another reason could be that the cluster environment causes galaxies to transform their properties, as they move from low density regions into the cluster core.  A number of plausible mechanisms have been proposed, including interactions with the ICM, interactions with the cluster gravitational potential and small scale galaxy-galaxy interactions \citep[see][and references therein]{Treu2003}. Each one of these mechanisms is expected to be effective in different regions of clusters and their outskirts, and to affect star formation in different ways and time-scales \citep[see e.g. the review by][]{Boselli2006}. 

For low-mass galaxies, however, the picture appears much clearer. Due to their low mass surface density and rotation velocity, dwarf galaxies are the most vulnerable systems to environment-related processes that quench star formation, such as the stripping of their halo and/or disk gas. \citet{Haines2007}, based on a large sample of dwarf galaxies, have found that passive dwarfs are  found only in very dense environments, typical of a cluster virial region, or as satellites of a more massive companion. Studies of three of the richest clusters in the local Universe, A2199 \citep{Haines2006}, Coma and A1367 \citep{Mahajan2010}, suggest that star formation in dwarf galaxies is quenched only in the center of the clusters, in contrast with massive galaxies that can become passive in all environments. All these evidences seem to imply that there exist fundamental differences in the evolution of giant and dwarf galaxies in clusters: while various physical mechanisms could be co-responsible for the evolution of massive galaxies, the evolution of dwarf galaxies seems to be primarily driven by the environment in which they are found.  

In accordance to the current cosmological paradigm \citep[e.g][]{Springel2005}, clusters at $z\sim 0$ have been found to accreate late-type galaxies along the filamentary structures that compose the cosmic web \citep[e.g.][]{Smith2011}. \citet{Mahajan2011} have found that post-starburst dwarfs are preferentially located  in the infall regions of Coma and A1367, suggesting that these galaxies experience a sudden quenching of star formation due to the interaction with the ICM. Moreover, before the gas gets totally stripped off, dwarf galaxies could experience an enhancement of star formation, either during their infall into the cluster along the filaments \citep{Mahajan2010,Porter2008}, or in the first stages of their encounter with the hot ICM \citep[pressure triggered star formation,][]{Abramson2011,Sun2007,Gavazzi2001}. Thus  SF dwarf galaxies are excellent probes to test the influence of the environment on the process of star formation and galaxy evolution. 

The gas interchange of a galaxy with the environment is observationally well confirmed. For example, the ram-pressure stripping \citep[RPS,][]{Gunn1972} can remove the interstellar medium (ISM) of a galaxy as it moves through the hot intracluster gas. Evidences of ongoing gas stripping, both in low and high luminosity galaxies, have turned out to be frequent in local Universe clusters, such as the Virgo cluster \citep{Fumagalli2011, Abramson2011,Kenney1999}, A1367 \citep{Gavazzi2001},  and A3627 \citep{Sun2007}. In the Coma cluster core several galaxies have been observed showing clear signs of gas stripping, such as \Ha or ultra-violet tails \citep{Yagi2010, Smith2010}. Additional observational evidences exist confirming the gas stripping in clusters, such as truncated ionized gas disks \citep{Koopmann2004, Koopmann2006, Cedres2009, Jaffe2011} and disturbed gas kinematics \citep{Jaffe2011}. Moreover, the atomic gas properties of cluster galaxies seem to tell the same story \citep{Haynes1984,Solanes2001,Chung2009,Levy2007}.

Simulations \citep[e.g.][]{Abadi1999,Mori2000,Tonnesen2007} suggest that RPS is efficient even up to the cluster virial radius and can remove completely the gas content in time-scales of $\sim10^9$ yr, comparable to the cluster crossing time.  Additionally, RPS has been found to be a multi-stage process \citep{Roediger2005} and recent simulations indicate that both varying ICM density \citep{Bruggen2008,Tecce2010} and inhomogeneities in the ISM \citep{Tonnesen2009} are expected to play an important role in the gas stripping effect. \citet{Tecce2011} report that the gradient of ram pressure becomes steeper with increasing cluster mass; in the massive clusters ($M\sim10^{15}M_\sun$), the ram pressure at the core is $\sim100$ times higher than at $R=R_{200}$\footnote{$R_{200}$ is the radius from the cluster center which defines a sphere with interior mean density 200 times the critical density and is used as an approximation of virial radius.}, whereas in galaxy group-sized haloes the ram pressure at $R_{200}$ is $\sim10\%$ of the central value.  \citet{Bekki2009} simulations suggest that even moderately strong ram pressure, e.g. in clusters of $M\sim10^{14}M_\sun$ or even in groups, could strip the hot gas halos of galaxies, an effect known as strangulation. In these simulations, the stripping of galactic halo gas seems to be more efficient than that of disc gas, and the efficiency increases with decreasing galaxy mass. 

Gas infall has also been proposed to take place in both low-mass irregular and high-mass spiral galaxies, in order to explain broad-band colors, gas fractions, SFRs, and metallicities at low and high redshifts \citep[see][and references therein]{Dalcanton2007}. Additionally,  interacting galaxies can undergo nuclear metal dilution due to gas inflows \citep{Kewley2010,Montuori2010,Rupke2010,MichelDansac2008}, resulting in the flattening of the gas-phase metallicity gradients and altering their positions on the mass-metallicity relation (MZR).

All these environmental effects described, on both the SFH and the gas content of a galaxy, would be expected to leave their imprint on galaxy metallicity.  Even the most recent events of star formation can enrich the gas with metals, so gas metallicities should be very sensitive to trace ongoing changes on the SFH and gas content \citep{Ellison2009}. SF galaxies, through their ionized gas emission, provide this observable information on nebular metallicity. Thus, studying the chemical abundances of SF galaxies in clusters can help appreciate the effect of the environment on the chemical evolution of cluster galaxies. 

Previous works \citep{Mouhcine2007,Cooper2008,Ellison2009} on the metallicity of SF galaxies as a function of the environment have not provided conclusive results, as  small variations of $\sim0.05$ dex have been derived. These works however, have been limited to the high stellar mass range, where there are not conclusive evidences whether the cluster environment could change significantly the SFH. Additionally, they have focused on the effect of local galaxy density, which possibly is not the only relevant parameter. 
Highlighting the cluster environment impact, HI deficient spirals in Virgo and Pegasus I clusters, have been found to be more chemically enriched than normal spirals \citep{Skillman1996,Robertson2011}, while this correlation between HI deficiency and metallicity has not been observed for field galaxies \citep{Robertson2011}. 

The gas metallicity of SF dwarf  galaxies in local Universe clusters has also been addressed. In Virgo dIrr, the gas metallicity has not been found to show a clear trend with the environment \citep{Vilchez1995, Lee2003, Vaduvescu2007}; however, \citet{Lee2003} have found that five of these dIrr are gas deficient with respect to field dIrr at comparable oxygen abundances, and this gas deficiency correlates with the X-Ray surface brightness of the ICM. Some SF dwarfs in Hydra cluster  \citep{Duc2001} have been found metal-rich for their luminosities and \citet{Vaduvescu2011}, comparing the MZR of SF dwarfs in Hydra, Fornax and Virgo, have suggested that differences in the MZR seem to exist for galaxies in such different environments. Going to a more massive cluster (A2151), \citet{Petr2011} have been able to observe dwarf galaxies, located in the cluster core, showing higher metallicities for their mass.

In the present study we extend these previous works, investigating the chemical history of SF dwarf galaxies in four clusters in the local Universe: Coma (A1656), A1367, A779, and A634. Coma and A1367  are among the nearest very rich galaxy clusters, and both belong to the large structure called the Coma supercluster. Recent insightful evidences on the assembly history and the SFH of the low-mass galaxy population in Coma supercluster \citep{Smith2011,Mahajan2011,Mahajan2010}  seem to indicate that it provides exemplary conditions for searching the potential imprints of the cluster environment on galaxy chemical enrichment. To investigate whether the mass of the host cluster could play a significant role in the chemical evolution of  SF dwarf galaxies, we include to our sample two lower-mass clusters: A779 and A634. In total, these  clusters are the four Abell clusters with SDSS spectroscopic data, that fulfill the criteria: i)  to be visible from the northern hemisphere ($\delta\gtrsim$ -25 deg), and ii) to be located at the same distance at $\sim100$ Mpc ($0.02<z<0.03$). Thus, the sample of clusters studied in the present work belong to a semi-spheric shell of the local Universe\footnote{In this semi-spheric shell are found three additional Abell clusters: A400, A539 and A2666, but they do not have SDSS spectroscopic data. Multifiber spectroscopic observations for these clusters will be presented in a forthcoming work (Petropoulou et al.,in prep).}, and span a wide range of halo masses from $10^{13}$ to $10^{15} M_\sun$.

We have used the latest spectroscopic release SDSS-III DR8, where new emission line measures have been provided, after correcting the spectra for the underlying stellar population. This is an important issue when deriving gas-phase metallicities, reducing biases that previous works could have been suffering. Additionally, three out of four  of the present sample  clusters were not included in previous releases than DR7. Anyway, they would have been excluded by the  previous studies on the gas-phase metallicity of SDSS cluster galaxies \citep{Ellison2009,Cooper2008,Mouhcine2007}, due to the redshift cut-off implemented in these works. Moreover, our analysis include,  apart from oxygen abundance, the N/O ratio, a relevant observable to appreciate the effects on cluster galaxies \citep[see also][]{Petr2011}.

This paper is organized as follows: In \S\ref{GAL} we describe the selection of the sample of SF dwarf galaxies in four nearby clusters, in \S\ref{DATA} we derive the spectroscopic properties of our galaxy sample and in \S\ref{MET} their gas-phase metal abundances. In \S\ref{MZR} we discuss the MZR of our  sample clusters and in \S\ref{hi} the  chemical enrichment of the SF dwarf galaxies relative to their HI mass content. In \S\ref{DIS} we discuss the environmental effects that could affect the chemical evolution of cluster galaxies and finally in \S\ref{SUM} we briefly summarize the findings of the  present work. In this work we adopt  the cosmological parameters $H_0$=73 km s$^{-1}$ Mpc$^{-1}$, $\Omega_\Lambda=0.73$, $\Omega_0=0.27$.

\section{THE GALAXY SAMPLE}\label{GAL}

In this paper we study the chemical properties of SF dwarf galaxies in the central region  (up to $\sim3R_{200}$) of four nearby clusters (Coma, A1367, A779, and A634), using spectroscopic data of SDSS-III DR8. In Table \ref{tbl1} we give details on the cluster properties and in Table \ref{tbl2} we give details on the regions studied in this work (see the captions of both Tables). 

Coma is a very massive cluster, with very high velocity dispersion and X-ray luminosity \citep{Ledlow2003}. There are two central cD galaxies NGC 4874 and 4889, and a subcluster, projected to $\sim1.5$ Mpc to the south-west of the core, and centered on NGC 4839, appears to be merging with the main cluster \citep{Briel1992,Neumann2003}. A1367 reveals a complex dynamical state \citep{Cortese2004}, with an elongated X-ray emission, showing multiple clumps, supporting a multiple merger scenario \citep{Donnelly1998,Sun2002}. Additionally, substructures falling into the cluster core \citep{Cortese2006,Gavazzi2003} suggest an early stage of formation of the cluster. \citet{Cortese2008a}, based on UV photometry,  have found  that star formation in Coma is substantially suppressed compared to that in the field, while A1367 has an abundance of bright SF galaxies \citep[see][]{Iglesias2002}. The optical luminosity function of galaxies has a much steeper faint-end slope in Coma than in A1367 \citep{Iglesias2003}.  On the other hand, A779 and A634 are two clusters of very low velocity dispersion. A634 X-ray luminosity is at the ROSAT detection limits, while A779 presents a nearly circular X-ray emission around the cD galaxy NGC 2832.

\input{tbl1}
\input{tbl2}

Cluster galaxies were selected using spectroscopic redshift information. In Figure \ref{fig1} we plot the velocity histograms of all galaxies with SDSS spectroscopic data in the areas considered. We perform a gauss fit to the velocity distribution of each cluster. The central velocity obtained by the fit is in good agreement with the mean cluster velocity given by NED, indicated in Figure \ref{fig1}. We consider as cluster galaxies all the galaxies with velocities $\mathrm{v}_\mathrm{clus}\pm3\sigma_\mathrm{v}$, where $\mathrm{v}_\mathrm{clus}$ is the mean cluster velocity and $\sigma_\mathrm{v}$ is the dispersion given by the fit. In Table \ref{tbl2} we give the velocity range considered for each cluster, and the total number of galaxies with SDSS spectroscopic data found within the area considered and within the respective velocity range.

To select the dwarf galaxies, as the four clusters are located at similar distances (m-M$\sim$35.0), we adopt the criterion used by \citet{Mahajan2010,Mahajan2011}: M$_{z}$$\geq$15, which results in a selection of $\sim(\mathrm{M}^\star+2) < \mathrm{M}_z< \sim(\mathrm{M}^\star +4)$, assuming $\mathrm{M}^\star_z=-22.32$ \citep[from][]{Blanton2001} and the SDSS upper magnitude limit r = 17.77. This criterion yields to a sample of galaxies of log mass range $\sim8.0-10.0$ M$_\sun$.

From this sample we select the SF dwarf galaxies adopting the criteria: SDSS BPT class=1 \citep[classification as star-forming, based on the BPT diagram,][]{BPT81}, and signal to noise (S/N)$>3$ for the lines \OIII5007, \NII6584, H$\alpha$, H$\beta$, \SII6717, \SII6731.  Finally, we have taken special care to include to  our final sample of  SF dwarf galaxies one spectrum per galaxy (a few galaxies have multiple observations), as well as to exclude spectra that correspond to distinct \HII regions of some parts of galaxies, because these spectra provide substantially underestimated mass values (see \S\ref{DATA}). The number of  SF dwarf galaxies for each cluster is given in Table \ref{tbl2}. For these galaxies we calculate the distance $R$ from the cluster center given in units of the $R_{200}$. In Table \ref{tbl2} we also give the number of  SF dwarf galaxies located in the cluster core, at distance $R\le R_{200}$, and the total number of dwarf galaxies at $R\le R_{200}$ for each cluster.

We note that there are more  SF dwarf galaxies, showing low S/N emission lines (the galaxies with BPT class=2 in SDSS), in the areas of the clusters of our sample, specially of Coma.  These galaxies could be suffering the quenching effect of the cluster environment, illustrated well in studies based on the \Ha EWs  of SF galaxies \citep[e.g.][]{Balogh2004,Rines2005}. In the present work, given that a high S/N emission line spectrum is required to derive properly spectroscopic properties (e.g. reddening coefficients) and gas metallicities, we do not include in our galaxy sample the low S/N SF population (we call them class 2 population). By doing this, we probably consider  the more recently accreted galaxies to the cluster environment, and as a consequence the potential suppressing of their star-formation could be observed in an early stage. As we will discuss in \S\ref{MZR}, the behavior of class 2 population would not change the conclusions of this work.

\citet{Poggianti2006}, based on a large sample of clusters, sampling the whole mass range from groups to massive clusters, have been able to show that the fraction of SF galaxies depends on galaxy mass, both for clusters at high-z and low-z. These authors found that about 20\% of the galaxies in clusters at $z\simeq0$ with $\sigma>500$ km s$^{-1}$ are SF. Due to the SN restrictions, and the SDSS spectroscopic data incompleteness \citep[specially in clusters, due to constraints in the fiber placement, see][]{Blanton2005a,Blanton2005b}, our galaxy sample is not a complete sample of the SF galaxies in these clusters. Thus, we should not compare our data with the rates found by \citet{Poggianti2006}. However, the goal of this work is to investigate the imprint of the cluster environment on the chemical evolution of SF galaxies in clusters, the general quenching effect in clusters being well established by several previous works. 

Figure \ref{fig2} shows the color g-i histogram of dwarf galaxies (green line),  SF dwarf galaxies (blue line), and  SF dwarf galaxies to $R\le R_{200}$ for each cluster (note that logarithmic scale has been used for Coma). It is well illustrated that the dwarf galaxy population in these clusters is composed by two main populations, red and blue, producing two maxima in the color histogram at g-i$\sim0.6$ and g-i$\sim1.1$, following the general bimodal distribution of the galaxy population \citep{Strateva2001, Baldry2004}.  Coma, has a larger fraction of red-sequence over blue SF dwarfs, in agreement with its low fraction of spiral galaxies  \citep{Dressler1980}. The dynamically younger cluster A1367, in turn,  hosts an almost equivalent blue and red dwarf population, as well as a larger number of SF dwarfs inside $R_{200}$. These differences are in agreement with previous findings on the color-magnitude diagrams of these clusters \citep[e.g.][]{Mouhcine2011,Terlevich2001}. We note that in all clusters there are some SF galaxies with red colors and that the SF dwarfs at $R\le R_{200}$ span the whole color range. We have performed Kolmogorov-Smirnov (KS) tests which have shown that the  SF dwarf galaxies to distances $R\le R_{200}$ and those to distances $R_{200}<R<3R_{200}$ are  statistically indistinguishable with respect to their $g-i$ color for Coma, with a probability of 89\%. Smaller probabilities have been found for the three remaining clusters (for A779 and A634 possibly due to the lower galaxy number).

\citet{Kewley2005} argued that using the 3-arcsec diameter aperture of the SDSS spectra to estimate SFRs and metallicities of late type galaxies located at $z< 0.04$ could introduce systematic biases. This is due to the significant radial metallicity gradients that can be found in SF spiral galaxies \citep[e.g][]{Kennicutt2003,Magrini2007,Bresolin2009}. However, in this work we focus our attention to dwarf/irregular galaxies, where spatial metallicity gradients are not expected to be important \citep{Kobulnicky1997,vanZee2006,vZH2006}. Thus, taking into account the smaller diameter of dwarfs, we do not expect aperture biases to be important for the galaxies of the present sample.

\input{fig1}
\input{fig2}

\section{SPECTROSCOPIC PROPERTIES}\label{DATA}

SDSS-III DR8 provides new emission line measures \citep{Tremonti2004,Brinchmann2004}, derived after correcting the spectra for the underlying stellar continuum, using the high-resolution population synthesis models of \citet{Bruzual2003}. The Balmer emission lines can be severely affected by the absorption of the underlying stellar component \citep{MartinManjon2008,MartinManjon2010} and this would affect the derived values of the reddening coefficient, the line indices and, as a consequence the gas-phase metallicity (see \S\ref{MET}). This effect is expected to be negligible here, allowing a reliable study of the galaxies' gas metallicity. We find that DR8 measures of Balmer emission lines such as H$\gamma$ and H$\delta$  are significantly improved as compared to DR7 spectroscopic data. 

We use the Balmer emission lines H$\alpha$, H$\beta$, H$\gamma$ (and where the S/N permits  also H$\delta$) to derive the reddening coefficients c(H$\beta$) for our sample of SF galaxies, using Case B approximation \citep{Osterbrock1989}, adopting the \citet{Cardelli1989} extinction law, and taking into account the line measurements error. \citet{Mahajan2010} provide Spitzer MIPS 24$\mu$m flux for a sample of Coma cluster galaxies. From this sample, 25 galaxies belong to our sample of SF dwarfs. For these 25 galaxies we find that the derived reddening coefficient  c(H$\beta)$ shows a tight correlation with their flux at 24$\micron$, in the line of \citet{Relano2010}, who suggest that the dust responsible for the Balmer extinction should be emitting at 24 $\micron$.

The presence of both singly and doubly ionized oxygen line transitions in the optical wavelengths has permitted to develop an efficient metallicity calibration \citep{Pagel1979} based on the indicator R$_{23}$=(\OII3727+\OIII4959+\OIII5007)/H$\beta$. Obtaining the \OII3727 line measures from DR8, however, resulted problematic. Although some spectra present high S/N \OII3727 line, no measure is provided by DR8. For the galaxies that SDSS does not provide \OII3726,3729 measures with S/N$> 3$, but \OII3727 is detectable in the spectrum with S/N$> 3$, we measure the \OII3727 integrated flux. We have verified that our measures and DR8 give consistent emission line fluxes for several emission lines (e.g. \NII/\Ha and \NII/\OII, when there is \OII).  
Table \ref{tbl3} presents the number of galaxies for which we have SDSS \OII3726,3729 measures, the number of galaxies for which we measure integrated \OII3727, the total number of galaxies with \OII measure and the total number of  SF dwarf galaxies for A1656 and A1367.
\input{tbl3}

Despite this effort, we could not obtain \OII3727 for all galaxies belonging to the four clusters of z$\sim$0.023. This is because \OII3727, redshifted to this velocity, lie at the edge of the wavelength range covered by SDSS spectroscopy. We have performed Kolmogorov-Smirnov (KS) tests and have verified that the subsample of galaxies with \OII measured (by DR8 or this work) and the whole sample of  SF dwarf galaxies are statistically indistinguishable in their properties: $M_z$, metallicity (by PMC09, see \S\ref{MET}) and mass. We conclude that the lack of \OII measures is a random effect, due to the radial velocity of each galaxy for the cluster velocity dispersion, and does not correlate with any of the fundamental galaxy properties.

We have used the reddening corrected line fluxes for our sample of galaxies to compute five standard optical line ratios, that we show in Figure \ref{fig3}, combined into three commonly used diagnostic diagrams. Our sample galaxies show typical line ratios of normal SF dwarf galaxies. Additionally, the galaxies located at $R\le R_{200}$ span the whole range of values in all diagrams, not showing any peculiarity with respect to the spectroscopic properties of their ionized gas.  Figure \ref{fig4} shows the excitation as measured by \OIII/H$\beta$ versus EW(H$\alpha$) (that we derived from the stellar absorption corrected \Ha flux and the adjacent continuum as provided in DR8) for the  SF dwarf galaxies in Coma (left) and A1367 (right). Filled symbols represent galaxies to distances $R\le R_{200}$ and again, we see that the galaxies in the cluster core are not different from the rest regarding excitation. 

\input{fig3}
\input{fig4}

We have also obtained the median estimate of the total stellar mass of the galaxies from the MPA-JHU spectroscopic catalog \citep{Kauffmann2003}. We have found that this galaxy mass estimate is consistent with the mass we derived using the k-correct algorithm \citep{Blanton2007}. We also have found a good agreement between the MPA-JHU mass estimates and \citet{Mouhcine2011} galaxy masses, derived by K band imaging, for 7 galaxies of our sample that belong also to the sample of \citet{Mouhcine2011}.

\section{OXYGEN ABUNDANCES}\label{MET}

Accurate abundance measurements require the determination of the electron temperature, which is usually obtained from the ratios of auroral to nebular line intensities such as \OIII4363 over \OIII5007. This is often referred to as the ``direct'' method. When auroral lines are not detected, empirical methods are generally used, based on the ratios of bright forbidden lines to hydrogen recombination lines. Among these, the most widely used abundance indicator is the R$_{23}$\footnote{R$_{23}$=(\OII3727+\OIII4959+\OIII5007)/H$\beta$} index \citep[although still carrying significant hazards, e.g. see][]{PMD2005}. An other extensively used parameter to derive gas phase metallicities is N2\footnote{N2=log(\NII6584/H$\alpha$)}  \citep[][PMC09 from now]{Storchi1994,PMC2009}. The use of this parameter has two important advantages: the relation between N2 and the oxygen abundance is single-valued and the emission lines involved are very close in wavelength, so the N2 parameter is almost free of uncertainties introduced by reddening correction or flux calibration. A third very useful index is O3N2\footnote{O3N2=log\{(\OIII5007/H$\beta$)/(\NII6583/H$\alpha$)\}} \citep{Alloin1979}, which shows a relatively tight and linear relationship with 12+log(O/H) \citep[from now on PP04]{Pettini2004} for O3N2$\lesssim2$.

The calibration of these parameters can be empirical \citep[e.g.][from now P10]{Pettini2004, Pilyugin2005, Pilyugin2010}, that is using direct measurements of the oxygen abundance of \HII regions in the Local Universe, or theoretical \citep[e.g.][]{McGaugh1991,Kewley2002,Tremonti2004,Nagao2006,Dors2011}, that is using photoionization models covering different ranges of physical parameters. Though, it has long been reported \citep[e.g.][]{Kewley2008} that there exist systematic differences that can reach up to $\sim0.5$ dex between most model and empirical calibrations, specially in the higher metallicity regime \citep[but see][]{Dors2011}.

A strong hint that the direct measures and the empirical calibrations should yield metallicities closer to reality was given by \citet{Bresolin2009}. These authors derived the metallicities of 28 \HII regions in the galaxy NGC300 using the direct method, and found that these metallicities agree with the abundances derived from young massive stars, with stellar and nebular abundances giving virtually coincident slopes and intercepts of radial gradients. An important quantity that should be taken into account when comparing stellar and nebular oxygen abundances is the amount of oxygen depleted onto dust grains in ionized nebulae. According to \citet{Peimbert2010} this fraction amounts to about 0.10 dex for low-mass galaxies. \citet{Bresolin2009} discussed this  effect and found that, even considering a gas depletion factor of $-0.1$ dex, the intercepts of the nebular and stellar abundance gradients would be consistent. Then they compared their direct estimates of the metallicity of the ionized gas with the most widely used model calibrations, and they report that model calibrations yield higher metallicities of about a factor of two in the O/H range derived. On the contrary, the abundances derived using empirical calibrations use to be in good agreement with their direct measurements. Another strong hint was given recently by \citet{Dors2011}. These authors produced models that reproduce O/H estimates consistently with the values derived using the direct method, for the upper range of O/H values, where all previous models showed systematic discrepancies.

\citet{SimonDiaz2011} derived the O/H abundance ratio using 13 B type stars from the Orion star forming region and obtained an excellent agreement with the O/H recombination line abundances derived for the Orion nebula by \citet{Esteban2004}. For the Orion nebula \citet{SimonDiaz2011} considered a correction of $\sim$0.12 dex due to the dust grain depletion, that is in excellent agreement with the value estimated by \citet{MesaDelgado2009}. Since recombination lines are generally very weak they are not observed in our sample of galaxies. In any case, this
analysis is based on the relative comparison of galaxy metallicities; thus no relevant effect is expected from the derivation of absolute abundances.

In this work we use the empirical calibrations of P10, N2 calibration of PMC09, and O3N2 calibration of PP04 to derive the oxygen abundances. We note that for our sample of galaxies always O3N2 $<2$, lying in the valid range of PP04 calibration. In Figure \ref{fig5} (left) we compare the oxygen abundances derived using the PMC09 and PP04 calibrations for the galaxies in the two most populated clusters of our sample, and we can see a very good agreement. The use of the P10 empirical calibration requires the measure of the \OII3727 line, that we do not  have for all our sample galaxies, see \S\ref{DATA}. In Figure \ref{fig5} (middle) we compare the oxygen abundances derived using the P10 and PMC09 calibration, for those galaxies with \OII measured. We see that the PMC09 calibration yields slightly higher metallicities ($\sim0.1$dex) than P10 calibration, in agreement with our previous considerations in \citet{Petr2011}. We also derive the oxygen abundance of galaxies with \OII measures using the model of \citet{Dors2011}, based on the diagnostic diagram \OIII/\OII versus \NII/\OII\footnote{We here adopt the value 12+log(O/H)$_\sun$=8.69 for the solar oxygen abundance \citep{Asplund2009}.}. We find that these models systematically underestimate metallicities for 12+log(O/H)$<$8.2. On the contrary, for 12+log(O/H)$\gtrsim$8.2, we find a good agreement with the empirical PIL10 estimates (see Figure \ref{fig5}, right). 
\input{fig5}

In order to avoid limiting our sample only to galaxies with \OII measures, in the following we use the PMC09 calibration of  the galaxy oxygen abundance. We are aware that PMC09 could slightly overestimate ($\sim 0.1$dex) the galaxy O/H abundance, but as discussed later on, we do not expect this to affect our discussion, as we always perform careful comparison of metallicities derived using the same method. We also use the N2S2 calibration of PMC09 to derive N/O ratios for our  SF dwarf galaxies as in \citet{Petr2011}.

In order to illustrate the properties of our galaxies, in Figure \ref{fig6} we plot the \Ha flux of the SF region covered by the SDSS fibre, for the SF dwarf galaxies in Coma (triangles) and A1367 (diamonds), versus the EW(H$\alpha$). The \Ha fluxes have been corrected for extinction (from the derived c(H$\beta$), see \S\ref{DATA}), and have been transformed to an equivalent number of ionizing photons Q(H), under the assumption that no ionizing photons escape the galaxies \citep{Osterbrock1989}. The points are color coded to the derived galaxy metallicity. Then we add the loci of Starburst99 models \citep{Leitherer1999} for instantaneous bursts of total mass $M_\star=10^7 M_\sun$ (continuous lines) and $M_\star=10^6 M_\sun$ (dashed lines) in massive stars for metallicities Z=0.02 (in red), Z=0.008 (green), and Z=0.004 (black). According to these models, and assuming single bursts, the SF regions studied would span a mass range of  $\sim10^6-10^7 M_\sun$.
\input{fig6}

\section{THE MASS-METALLICITY RELATION}\label{MZR}

In Figure \ref{fig7} we plot the galaxy metallicity, as measured by 12+log(O/H), versus galaxy stellar mass, for the  SF dwarf galaxies in Coma (A1656, triangles), A1367 (stars), A779 (squares), and A634 (circles).  Filled symbols correspond to galaxies to a distance $R \le R_{200}$ from the cluster center. We see that the  SF dwarf galaxies follow well defined sequences on these plots. Having a look at the MZR of A1656 and A1367, we observe that at low galaxy masses ($10^8 M_\sun<M_\star<10^9 M_\sun$), the derived metallicities cover a range of $\sim0.4$ dex, while at higher  masses ($10^9 M_\sun<M_\star<10^{10} M_\sun$) the relation becomes tighter, appearing to shape a triangle instead of a simple linear correlation. Additionally, we observe that, for the same bin of mass, galaxies at $R \le R_{200}$ are preferentially located at the upper part of the global sequences. For the two clusters (A779, A634) of lower mass, we do not verify neither of these two features, as we discuss in the following.

We first note that the behavior observed seems independent of the abundance calibration used. When we use the PP04 calibration of O3N2, we obtain the same mass-metallicity sequences and the  SF dwarf galaxies in the core of A1656 and A1367 still crowd the upper part of the MZR (we can not do the same using P10 calibration because we do not have enough galaxies with measured \OII inside $R_{200}$). As a consequence, the effect observed seems to be significant, despite the intrinsic errors of the N2 and O3N2 calibrations. 

Second, in \S\ref{DATA} we have verified that our sample galaxies do not show any systematic differences regarding excitation, as a function of their location in the cluster (in or out $R_{200}$). 
Thus, the higher metallicities derived for galaxies at $R \le R_{200}$, with higher \NII/\Ha, are not biased by excitation effects \citep[see][for a thorough discussion on this effect]{Berg2011}. 

The third sanity check has been to test whether the position of the SF dwarfs at $R \le R_{200}$ on the  MZR, is the result of some bias in the mass estimate.  We have found that galaxy stellar mass, versus other galaxy properties, such as broad-band colors and excitation, do not show any correlated dependance of the cluster-centric distance of the galaxies. After all these tests, we assume that the shift observed toward higher metallicities, for galaxies at $R \le R_{200}$, mainly in Coma, should not be the result of some bias in the abundance and/or mass estimates.  

In order to explore further the trends that appear in Figure \ref{fig7}, we have performed a bisector linear fit to the MZR of each cluster, considering first the galaxies at  $R \le R_{200}$, second the galaxies at $R > R_{200}$, and third all the galaxies together. In Table \ref{tbl4} we present the obtained slopes and the corresponding errors from the dispersion around the linear fit. We observe that the MZR fits for galaxies at $R>R_{200}$ have slopes $\sim 0.3$ for all the clusters of our sample.  \citet{Lee2006} have derived the MZR for 25 nearby dwarf irregular galaxies ($10^6<M<10^9 M_\sun$), extending the well-known SDSS MZR \citep{Tremonti2004} to the lower mass range. \citet{Lee2006} have found a very tight correlation, over the whole stellar mass range, with slope $\sim0.3$; thus the slopes derived here  are in very good agreement with their results.
As we discuss further in \S\ref{DIS} this value of the slope is supported by the semi-analytic models and hydrodymanic simulations of \citet{Finlator2008} and \citet{Dave2011} where momentum-driven wind scalings are introduced to explain the MZR.
\input{tbl4}

For Coma, and to less extend for A1367, a flattening of the MZR is observed when considering galaxies inside $R_{200}$, with the fits showing clearly smaller slopes, within the quoted errors  (see Table \ref{tbl4}). This is made evident in Figure \ref{fig7}, where we plot with a continuous line the linear fit for galaxies at $R\le R_{200}$ and with dashed line the fit corresponding to galaxies at $R>R_{200}$.\footnote{We note here that the class 2 population refered to in \S\ref{GAL}, span the higher mass range, where we have seen that the MZR becomes tighter, and had it been included, would not have significantly changed the derived slopes.} 
For A779 and A634, considering the large errors (due to the reduced number of SF dwarf galaxies in these clusters), we do not appreciate any difference in the slope inside and outside the cluster core. In Figure \ref{fig7} we plot the overall fit to the MZR for A779 and A634.

In Figure \ref{fig8} we plot the N/O ratio versus galaxy stellar mass, and again we find a good correlation, even tighter than the MZR, for all our  clusters. Again, filled symbols correspond to galaxies to a distance $R \le R_{200}$ from the cluster center. As before, galaxies in the cluster core of Coma and A1367 tend to crowd the upper part of the global sequence. Oxygen, produced in Type II SNe, is typically released after $\sim$10 Myr, while the bulk of nitrogen is produced and released over a substantially longer period, $\gtrsim$ 250 Myr. This is an important piece of information to take into account when one attempts to identify the mechanisms relevant for the chemical evolution of cluster galaxies.  As we will discuss further in \S\ref{DIS}, this chemical ``clock'' seems to imply that if there is an environmental effect driving the observed difference in the abundance ratio, this should be acting since at least $10^8$ yr ago. 

\input{fig7}
\input{fig8}

Observing the triangular shape of the MZR of A1656 and A1367 in Figure \ref{fig7}, we investigate whether there could be a physical cause for the galaxies showing lower metallicity for the same bin of mass. The MZR has been found to show a second parameter dependence on SFR \citep{Amorin2010, Mannucci2010, Lara2010, Cresci2011}. This can be easily understood by the correlated behavior of gas-phase metallicity and SFR after a gas infall event: when a galaxy accreates metal-poor gas, its metallicity is diluted to values typical of lower mass galaxies, while its final mass increases, and concequently the galaxy moves below the MZR. In the same time, the presence of large amount of gas stimulates star-formation, and this is the reason why galaxies with higher SFR  have lower metallicities at a given stellar mass. This should be a transient phase, and when the galaxy will consume the gas, producing the new metals, will return to the mean MZR \citep{Dave2011,Dalcanton2007}.

\citet{Mannucci2010} claimed that the MZR is the projection in the 2D space of a more fundamental 3D relation between stellar mass, gas metallicity  and SFR and proposed the quantity $\mu=\log M_\star-0.32\log$(SFR) which defines a projection of the MZR that minimizes the scatter of local galaxies. We have used the median estimate of the total SFR provided by DR8 \citep{Brinchmann2004} to explore whether the quantity $\mu$ could decrease the scatter observed in Figure \ref{fig7}. The SDSS SFR estimates have been derived by combining emission line measurements of the fibre spectrum and applying aperture corrections by fitting models to the photometry outside the fibre \citep[as in][]{Gallazzi2005, Salim2007}.  In Figure \ref{fig9} we plot the oxygen abundance 12+log(O/H) versus $\mu$ for the SF dwarfs in A1656 and A1367, and we see that the scatter drops for the galaxies at lower metallicities for the same bin of mass (the Spearman's correlation coefficient increases from 0.73 in  Figure \ref{fig7} to 0.77 in Figure \ref{fig9} for A1656, and from 0.80 to 0.83 for A1367). However, the separation we observe in the MZR between galaxies inside and outside $R_{200}$ is kept equally. 

\input{fig9}

Moreover, in this metal dilution scenario, the N/O ratio should not be affected \citep[e.g. the ``green pea'' galaxies show normal N/O ratios for their mass,][]{Amorin2010}, and the relation between N/O versus stellar mass should not show correlated scatter. However, as Figure \ref{fig8} shows, the N/O ratio segregates galaxies inside $R_{200}$, these galaxies also showing higher N/O ratios for the same bin of mass. The effect of infalling gas is expected to be more relevant for galaxies at high redshift \citep[e.g.][]{Cresci2010,Tacconi2010, Dekel2009}, thus the fundamental metallicity relation indroduced by \citet{Mannucci2010} considerably accounts for the evolution of the MZR with redshift \citep{Erb2006, Maiolino2008, Mannucci2009}. But in the local Universe, where our sample of clusters are located, different mechanisms should be invoked to explain the trends observed.

The question arises now as to whether the galaxies inside $R_{200}$ have evolved in a different way than the other cluster galaxies, rendering them chemically more enriched than the galaxies at $R>R_{200}$, or alternatively whether less metallic galaxies, for the same bin of mass, are more vulnerable to the quenching effect of the cluster environment, and this could be the reason we do not observe them preferentially in the cluster core. To investigate the second idea we searched whether there is any observable trend with the distance from the cluster center of the morphological type of our sample of SF galaxies.

We have calculated the standard concentration index $C=R90/R50$,  where $R90$ and $R50$ are the radii enclosing 90 and 50 per cent of the Petrosian r-band luminosity of the galaxy. We have found that our sample galaxies show typical $C$ values as related to late-type galaxies \citep[$\sim2.3-2.5$, see e.g][]{Shimasaku2001,Strateva2001} and we see no trend of the $C$ value in and out $R_{200}$. We have additionally checked that galaxies in the same bin of mass do not show any concentration-metallicity correlation. A more detailed morphological study could be of interest here; \citet{Penny2011} found remarkably smooth structures of dwarf galaxies in the Perseus cluster, and suggest that dwarfs in cluster cores should be highly dark matter dominated to prevent their tidal disruption by the cluster potential. In the present study we can not verify the hypothesis that SF galaxies inside $R_{200}$ are  more concentrated/compact, and consequently more resistent to the hostile cluster environment. Thus, morphology does not seem to explain their relative position in the upper part of the MZR; this should be the result of some different evolution.

To investigate whether the scatter observed in the MZR could be related to differences in the underlying stellar population, we have obtained for all our cluster galaxies the spectral index $\mathrm{D_n}$ (4000) \citep{Balogh1999}, after correction for emission lines, as given by SDSS DR8. All our SF dwarfs show $\mathrm{D_n} (4000) < 1.4$, which corresponds to typical ages $<1$ Gyr. Additionally, we have seen that the gas-phase metallicity does not show any correlated behavior  with the age of the underlying population. In \S\ref{DATA} we have shown that SF dwarfs  at $R\le R_{200}$ in Coma do not show any observable difference in the g-i color distribution as compared to the galaxies at $R>R_{200}$. Thus, the more metallic SF dwarf galaxies found in the cluster core, specially of Coma, are neither older nor redder than the rest of SF galaxies.

To quantify the effect observed in the MZR for Coma and A1367, in the upper panel of Figure \ref{fig10} we plot the mean difference of the derived 12+log(O/H) abundance for each galaxy with respect to the 12+log(O/H) given by the bisector linear fit, as a function of the cluster-centric radial distance, sampled in a bin of 0.5 $R_{200}$. The errors correspond to the standard deviation from the mean value of log(O/H)-log(O/H)$_\mathrm{fit}$. In the lower panel we plot the same for the log(N/O) ratio. We observe that in Coma, the mean difference in 12+log(O/H), in the closest bin to the cluster core, is positive and above the rms error (i.e. positive for all objects), and can reach above $\sim0.15$ dex. The same is found for the log(N/O) ratio, which can get up to $\sim0.27$ dex difference in the cluster core. In A1367 smaller differences are obtained, of $\sim0.05$ dex in log(O/H) and $\sim0.1$ dex in log(N/O). This divergence, combined to the fact that for the low-mass clusters of our sample, this trend has not been revealed at all, seem to indicate that the cluster mass is a most relevant parameter. As it will be discussed in \S\ref{DIS}, what seems to drive the disparate evolution of SF dwarf galaxies in Coma appears to be related to the properties of the forceful ICM of this cluster.

\input{fig10}

We explore the dependence of the trend observed for our cluster galaxies as a function of the local galaxy density, using the density estimator $\Sigma_{4,5}$.\footnote{Defined as the logarithm of the density obtained to the average of the projected distances to the fourth and fifth nearest neighbors, including all galaxies with SDSS spectra, within the cluster region \citep[see][]{Petr2011}.} In the upper panel of Figure \ref{fig11} we plot the mean difference of the derived 12+log(O/H) for each galaxy with respect to the 12+log(O/H) given by the bisector linear fit, as a function of $\Sigma_{4,5}$, in a bin of 0.5 dex. The errors correspond to the standard deviation from the mean value of log(O/H)-log(O/H)$_\mathrm{fit}$. In the lower panel we plot the same difference for the log(N/O) ratio. In the highest local density bin we find a mean difference in log(O/H) of $\sim0.05$ dex for both Coma and A1367, in agreement with the previous findings \citep[e.g.][]{Ellison2009}. Log(N/O) presents a measured difference of $\sim0.15$ dex for both clusters.

\input{fig11}

We note here that estimating the local galaxy density for clusters with large velocity dispersions  as in Coma ($\sigma_\mathrm{V}\sim$1000 km s$^{-1}$)  and A1367, bears significant hazards. Allowing neighbors to have velocity differences of the order of $\sigma_\mathrm{V}$,  could result in counting out galaxies in the cluster core, and thus underestimating density. Converselly, allowing velocity differences of 2$\sigma_\mathrm{V}$ or 3$\sigma_\mathrm{V}$  could introduce a severe bias to relatively isolated objects, overestimating their density. In Figure \ref{fig11} we use $\Sigma_{4,5}$ derived permitting neigbors to have velocity differences of 2000 km s$^{-1}$, but we have checked that the behavior does not change using velocity difference of $\sigma_\mathrm{V}$, $2\sigma_\mathrm{V}$, and $3\sigma_\mathrm{V}$.

Previous works have pointed out the effect of the environment on the gas-phase metallicity of galaxies \citep{Mouhcine2007,Cooper2008,Ellison2009} and have found that galaxies within dense environments show statistically higher metallicities of $\sim0.05$ dex at the same bin of galaxy mass. These works, however, have related this behavior to local galaxy density. Specially \citet{Ellison2009} have discussed the importance of cluster membership, and although they have found that enhanced metallicities are present at distances $R<R_{200}$  from the cluster center, they have concluded that the enhancement observed is driven by local overdensity and does not depend on cluster properties such as $R_{200}, \sigma_\mathrm{V}$,  or cluster mass.

There is a general good relation between the cluster-centric distance and the local galaxy density, within the cluster virial radius \citep[e.g.][]{Rines2005}, so the metallicity enhancement of the cluster core galaxies is expected to appear as a function of the local galaxy density as well. However, in the present work we find that the enhancement of the gas-phase metallicity is sensitive to the cluster mass, and appears to be more prominent if we consider the $R_{200}$ region of a massive cluster such as Coma. This evidence points towards a possible connection of the chemical enrichement of cluster galaxies with their ICM properties.

\section{CHEMICAL ENRICHMENT VS HI MASS}\label{hi}

In clusters, the gas interchange of a galaxy with its environment is expected to be very relevant. Observable gas tails give evidence of ongoing gas stripping  and a significant fraction of spirals have been found to have truncated ionized gas disks compared to their stellar disks (see \S\ref{INTRO}). Recently, \citet{Jaffe2011}, based in a large sample of emission line galaxies from the EDisCS\footnote{European Southern Observatory Distant Cluster Survey} sample at intermediate redshifts ($0.4<z<1$), found that the fraction of kinematically disturbed galaxies increases with cluster velocity dispersion and decreases with distance from the cluster centre, but remains constant with projected galaxy density. In addition, disturbed gas kinematics does not co-occur with morphological distortions as traced by optical (HST) imaging, suggesting the mechanism that affects most the gas of cluster galaxies has to be linked with the ICM.

The atomic gas is expected to be the first to suffer the ram-pressure stripping in the cluster environment. Indeed,  cluster spiral galaxies are more HI deficient than similar galaxies in the field, and this deficiency appears to be increasing towards the cluster center \citep{Haynes1984}. Additionally, because of the shallower gravitational potentials, dwarfs can be extremely fragile to RPS \citep{Solanes2001}. In the Virgo cluster core, \citet{Chung2009} found many HI deficient galaxies, and at intermediate distances, at $\sim$1Mpc from the center, they found a remarkable number of galaxies with long, one-sided HI tails pointing away from M87. Truncated HI disks were found even in Pegasus I cluster \citep{Levy2007}, a cluster with a low level of X-ray emission.

An interesting method to check for modulations of the HI content of SF galaxies is through comparing the galaxy chemical enrichment with the theoretical values predicted by the ``closed-box''  model \citep{Edmunds1990}. According to this model, the mass fraction of metals should be a direct function of the gas mass fraction: 
\[\mu=M_{gas}/(M_{gas}+M_{\star})\]
Here, the gas mass is considered to be the mass of HI ($M_\mathrm{HI}$) with a correction for neutral helium $M_{gas}=1.32M_\mathrm{HI}$. We neglect the contribution due to molecular hydrogen, as this seems to be small and difficult to evaluate. According to \citet{Israel1997}, for Magellanic type irregular galaxies, the $\mathrm{H_2}/\mathrm{HI}$  mass ratio is 0.2, which would mean an increase of 0.08 dex in the $M_{gas}$ estimate. However, the CO to $\mathrm{H}_2$ conversion relation is not well known for low metallicity systems and the $\mathrm{H}_2$ fraction might be even higher \citep[e.g.][]{Keres2003}.

Comparing with chemical evolution models is actually possible only when information on the HI mass  of a galaxy is available. Two of our sample clusters have recently published HI  data. \citet{Cortese2008b} have presented 21 cm HI line observations of the central part of A1367, as part of the Arecibo Galaxy Environment Survey (AGES). This sample covers $\sim20\%$ of the total region covered by our sample and $\sim70\%$ of the region with $R\le R_{200}$ and their  HI lower mass limit, at A1367 distance, is $6\times10^{8}$M$_{\sun}$.  They have detected 57 galaxies that belong to A1367 cluster, out of which 17 also belong to our sample of  SF dwarf galaxies.  \citet{Haynes2011} have recently released a catalog with 21 cm HI line sources, covering $\sim40\%$ of the final ALFALFA\footnote{Arecibo Legacy Fast Arecibo L-band Feed Array (ALFALFA).} survey area. This release includes $\sim50\%$ of the area of A1656 covered by the present work, also $\sim50\%$ of the area with $R\le R_{200}$, and their low HI mass limit at the Coma cluster distance, is $\sim 4\times 10^{8}$M$_{\sun}$  \citep[$M_\mathrm{HI}>10^7M_\sun$ at the Virgo cluster distance,][]{Giovanelli2005}. We have found 31 objects in common with our sample of  SF dwarf galaxies.\footnote{We have derived the $M_\mathrm{HI}$ using the standard formula $M_\mathrm{HI}=2.36\times 10^{5} D^2 F_\mathrm{HI}$ \citep{Haynes2011} and adopting the same mean distance $D$ for all cluster galaxies.}

In Figure \ref{fig12} (left) we plot the oxygen abundance versus the gas mass fraction (in the form $\log \ln 1/\mu$) for  A1656 (triangles) and A1367 (stars) for the galaxies we have HI data (filled symbols mean $R\le R_{200}$). In Figure \ref{fig12}  (right) we plot the gas mass fraction $\mu$ versus the stellar mass. We add (with smaller points) all the  SF dwarf galaxies of our sample that do not have HI measurements, but they are located  within the regions mapped by AGES and ALFALFA (up to the present release).  We assign to these galaxies the HI mass detection limit of each survey, this being an upper limit of the HI mass of these galaxies (the arrows indicate the direction to which these upper limits could be displaced). Again, open and filled small symbols mean outside and inside the cluster $R_{200}$ respectively.

\input{fig12}

We observe that A1656 dwarf galaxies located to distances $R\le R_{200}$ (filled triangles), all except one, have been assigned an upper limit. In contrast, inside $R_{200}$ of A1367 we have both upper limits and detections. This could be an evidence of the stronger and more effective ram-pressure exerted by the ICM of A1656 than that of A1367. \citet{Tecce2010} have found that at low-z, the mean ram-pressure is $\sim 10^{-11} h^2$ dyn cm$^{-2}$ in clusters with virial masses $\sim10^{15} M_\sun$, while in lower mass clusters of $\sim10^{14} M_\sun$, the mean ram pressure is approximately one order of magnitude lower. At distances $R>R_{200}$ (open triangles), there are measured HI masses as well as upper limits for both clusters. Theoretical models have started to investigate the effects of the ICM turbulence, substructure and shocks during groups infall, finding that RPS can be effective even  at the cluster outskirts and fillaments, as well as during tidal interactions \citep[e.g.][]{Tonnesen2008,Kapferer2008}.

We then compare the position of our galaxies in the left plot of Figure \ref{fig12},  with  model predictions for the yield. The green continuous line indicates the model value $y_o=0.0074$, that is the theoretical yield of oxygen expected for a Salpeter IMF and constant star formation rate, for  stars with rotation following \citet{MM2002} models \citep{vZH2006}. Both, infall and outflow of well-mixed material will result in effective yields that are less than the true yield, as the enriched material is either diluted or lost from the system. The blue dashed line corresponds to a lower yield $y_o=0.002$, almost 1/4 of the model prediction, which is the effective yield found for a fraction of the sample of isolated dwarf irregular (dI) galaxies of \citet{vZH2006}.

Some of the SF dwarf galaxies located at cluster-centric distances $R>R_{200}$ in both clusters, and a fraction of the SF dwarfs in A1367 cluster core ($R\le R_{200}$), appear to follow the theoretical closed-box model predictions (considering our metallicity typical uncertainty $\sim 0.1$ dex, see \S\ref{MET}). These galaxies in the core of A1367 could be ``newcomers''  \citep[see also][]{Petr2011}, observed before the action of RPS lowers significantly their atomic gas content, shifting them towards lower effective yield values. 

We also observe a tendency to lower effective yields as metallicity increases, suggesting that these galaxies host lower HI mass than  the expected  by the closed-box model \citep[a similar behavior was found for A2151 SF galaxies, see][Fig 14]{Petr2011}. The atomic gas deficiency is confirmed when we compare the HI content of our sample galaxies (measurements or upper limits) with the $M_\mathrm{HI}$  of field counterparts of the same absolute magnitude, as calculated following \citet{Toribio2011}. Two representative examples are the dwarf galaxies CGCG 97-073 and  CGCG 97-079 (marked on the plot) that have been investigated before \citep{Iglesias2002,Gavazzi2001}, showing long (H$\alpha$) tails that reveal their recent interaction with the ICM. The HI content of  CGCG 97-079 appears to be severely affected: this galaxy has the lower HI mass detected and lays below $y=0.002$ in the left panel of Figure \ref{fig12}. CGCG 97-073 in turn does not appear to have lost a large amount of HI yet.

We conclude that, to the distance of Coma and A1367, given the detection limit of the HI data, we observe the atomic gas of galaxies that still contain a substantial fraction of their HI mass. The almost absolute lack of detections in the central part of A1656, indicates that  galaxies there have been severely affected by the cluster environment, and the ram-pressure stripping has partly or completely removed their atomic gas content, rendering them undetectable in 21 cm HI line. In turn, in the central part of A1367  there are some SF dwarf  galaxies still with detectable HI mass (i.e. possible ``newcomers'').

\section{DISCUSSION}\label{DIS}

It is well established that dense environments, such as cluster cores, present larger fractions of passive galaxies than the field \citep[e.g.][]{Dressler1980, Poggianti1999, Treu2003, Finn2005, Poggianti2006}. This confirms the important influence of the environment on galaxy evolution. Although the big debate on nature versus nurture remains open, accumulated evidence seems to indicate that, in the particular case of dwarf galaxies, environment is a fundamental driver of their evolution \citep{Mahajan2010, Haines2007, Haines2006}. It has been found \citep{Smith2011,Porter2008} that significant infall of low-mass SF galaxies exists along the filamentary structures onto the densest clusters. Reaching cluster cores, dwarf galaxies can experience a star-burst event, induced by their first interaction with the ICM \citep{Gavazzi2001, Sun2007, Levy2007,Petr2011}. Then, RPS is expected to be very efficient \citep[e.g.][]{Tonnesen2007} and it could strip the gas of a dwarf galaxy in very short timescales ($\sim10^{9}$yr) converting a large fraction of dwarf galaxies, first into post-starburst as those observed in Coma \citep{Mahajan2011,Poggianti2004}, and finally into passive galaxies.

Despite the fateful ``switching off'' of star formation predicted by the models of RPS, we have found quite a number of  SF dwarf galaxies in our clusters. The selection of our galaxy sample to have strong emission lines means that they do still contain ionized gas. However,  as we show in \S\ref{hi}, a large fraction of these galaxies do not contain the amount of HI mass expected by the closed-box model, indicating that they should be suffering RPS which has first affected their neutral gas content. \citet{Tecce2010} simulations seem to indicate that at $z\sim0$, 50\% of galaxies inside the virial radius ($R_{vir}$) have experienced important ram-pressure effects. These pressures appear to have a strong effect on the cold gas content:  70\% of the simulated galaxies within   $R_{vir}$ have been found to be completely depleted of cold gas. Additionally, \citet{Tecce2010} models have suggested that the rate at which the cold gas of a galaxy is stripped, depends on the halo virial mass of the host cluster; thus less massive galaxies within massive clusters are the most affected.

In order to investigate the chemical evolution of the  SF dwarf galaxies in the present  sample  of clusters, we have derived their MZR. The MZR of galaxies is well established \citep[e.g.][]{Tremonti2004, Lee2006}. There exist a variety of chemical evolution models and hydrodynamical simulations advocating different physical mechanisms to explain this relation. For example, the ejection of metal-enriched gas by galactic outflows, triggered by e.g. (multiple) supernovae explosions, could be more efficient in systems with shallower potential wells \citep[e.g.][]{Larson1974, Marconi1994, DeLucia2004}. A variable efficiency of star formation, increasing with galactic mass (often named ``downsizing''), could also be invoked to reproduce the MZR  \citep[e.g.][]{Matteucci1994, Tissera2005, DeRossi2007, Calura2009}. For dwarf galaxies, where supernova winds are expected to be more important, a combination of downsizing and winds could affect the shape and slope of the MZR \citep{Tassis2008, Spitoni2010}. Other mechanisms, such as dilution caused by infall and variations in the IMF have also been proposed, however they show problems reconciling dwarf SFRs and metallicities \citep[see the review by][]{Tolstoy2009}.

Recently it has been found that hydrodynamic simulations that incorporate momentum-driven wind scalings provide among the most successful overall fits to a wide range of observed galaxy properties \citep{Finlator2008,Oppenheimer2010,Dave2011}. \citet{Finlator2008} developed a simple analytic model to understand the MZR, where the gas-phase metallicity of a galaxy is set by a balance between inflow and outflow, plus star formation. In this ``equilibrium'' model the mass outflow rate ($\eta$, i.e. the balance between inflow and outflow) scales inversely with circular velocity, and this naturally reproduces a slope of the MZR $Z\varpropto M_\star^{0.3}$ at $M_\star<10^{10.5}M_\sun$. \citet{Lee2006} observationally obtained this slope of the MZR for a sample of nearby dIrr galaxies, with a remarkably small dispersion. This value of the slope is in agreement with the slope given by the linear fit of our sample galaxies, located either in the low-mass clusters A779 and A634, or at the outer regions of Coma and A1367.

\citet{Dave2011} discuss the effect of SFR and of environment on the MZR, and find that the overall trends are consistent with the expectations from the ``equilibrium'' model. The scatter observed in the MZR for our sample galaxies does not correlate with SFR (\S\ref{MZR}), thus it does not seem to be related to metal dilution suffered by the galaxies showing lower metallicities at the same bin of mass. Regarding environment, \citet{Dave2011} hydrodynamic simulations reproduce low mass galaxies in dense environments showing higher metallicities for the same bin of mass than galaxies at lower densities. The authors suggest that this should not be the result of galaxies processing more gas into stars, but rather the consequence of the environmental dependence of wind recycling.  Much of the material entering into the galaxies' ISM at z=0 is recycled winds \citep{Oppenheimer2010}, that is the reaccretion of the wind material when the wind velocity does not exceed the galaxy escape velocity \citep[see also][]{Bekki2009a}. This mechanism accounts for the fact that metallicity differences seem to disappear for high mass galaxies: at high galaxy mass, wind recycling is so effective that metallicity approaches the theoretical yield regardless of environment. Instead, at low masses, most ejected material could normally escape the galaxy \citep{Oppenheimer2010}, unless the galaxy resides in a dense environment, with a hot gaseous halo that could significantly slow winds.

In \S\ref{MZR} we have seen that SF dwarf galaxies in the Coma cluster core, and to less extend in A1367, show higher metallicities for the same bin of mass than galaxies outside $R_{200}$. This results in the flattening of the MZR for dwarfs to distances $R\le R_{200}$. The amount of this flattening appears to be related to cluster mass, with the less massive clusters of our sample not showing at all this behavior. This observed trend matches well the above wind recycling scenario. Models suggest that at high density environments (i.e. in cluster cores) galactic winds can be suppressed by the high pressure of the ICM \citep{Schindler2005,Kapferer2006,Kapferer2009}. Accordingly this could cause faster recycling and thus preventing galaxies from loosing their metals. The suppression to take place needs ICM pressures $\gtrsim10^{-12}$ dyn cm$^{-2}$ \citep{Schindler2008}. \citet{Tecce2011} have found that on the outskirts of clusters as (or more) massive as Coma ($>10^{15} M_\sun$) the ram-pressure reaches $\sim 5 \times 10^{-12}h^2$ dyn cm$^{-2}$, the same order of magnitude as in the core of clusters of virial mass $\sim10^{14} M_\sun$. In turn, the ram-pressure in the core of massive clusters is expected to be $\sim 100$ times higher. Consequently, as we go to clusters of higher X-ray luminosity,  higher would be the pressure exerted and the wind suppression is expected to be more effective. The wind suppression in the cluster cores could approximate the no-wind case depicted at \citet{Dave2011} hydrodynamic simulations, where the slope of the MZR is found to be flatter than when considering wind scalings.

The important role of the density of the ICM can be supported by \citet{Poggianti2009a,Poggianti2009b} findings. These authors have found that the fraction of post-starburst galaxies depends on cluster mass, and the fraction of  spiral galaxies in clusters in the Local Universe is anticorrelated with $L_X$ (whereas no trend is observed with cluster velocity dispersion).  
Now it seems to turn up that it is not only to the quenching efficiency, but also to the chemical evolution of low-mass cluster galaxies that the properties of the ICM could play an important role.

Another possibility that emerges from the ``equilibrium'' model scenario is that galaxies in denser regions could result having higher metallicities as a concequence of curtailed  inflow, due to gas stripping and/or strangulation. The relation $Z\varpropto M_\star^{0.3}$ is the result of the balance between inflow and outflow processes. If gas accretion is truncated by some sort of environmental suppression, then the galaxy consume its gas to form metals along the locus of $Z\varpropto M_\star$ and will move above the mean MZR  \citep{Dave2011}. If accretion does not restarts, the galaxy will end up exhausting all the available gas for star-formation and finally be transformed to a passive galaxy. Accordingly, our  SF dwarf galaxies that lie above the MZR could be in this particular phase of environmental quenching, where enhanced metal enrichment precedes the switching off of star-formation. As we show in  \S\ref{hi} the majority of  galaxies inside $R_{200}$ seem to have suffered important gas removal, specially in A1656.

The above picture could describe well what is happening to the galaxies in the Coma cluster core, where SF dwarfs show on average $\sim0.15$ dex higher metallicities than the overall MZR. By reaching the outskirts of this massive cluster, the ISM-ICM interaction can produce star-burst events, that accelerate gas depletion in cluster dwarf galaxies, as compared to their isolated counterparts. In the same time, galactic winds should have started to get suppressed, preventing metal lost from the low-mass galaxies, that otherwise would have been expected to be very efficient. As RPS gets shearing the gas content (starting from the halo gas reservoir and the HI disk), the infall of pristine gas gets truncated, and the chemical enrichment follows a steeper path on the mass-metallicity plane. By the end of their star formation, the low-mass galaxies could have experienced a significant metallicity enhancement. Considering that this shutting-off in Coma could take  $\sim10^9$ yr (the crossing-time of this cluster), there is enough time also for nitrogen to get delivered to the ISM, yielding the observed trend in the N/O ratio. In Figure \ref{fig13} we give a schematic representation of this scenario.

\input{fig13}

Moreover, low-mass galaxies could reach clusters already chemically enhanced up to some degree. Recent works \citep{Mahajan2010, Porter2008} have found an increased star formation in galaxies falling into clusters along supercluster-scale filaments. Additionally,  simulations \citep[e.g][]{Bekki2009} suggest that even moderately strong ram pressure, such as  in group-like environments, could strip the hot gas halos of galaxies, with efficiency increasing at the low-mass regimes.  Consequently, low-mass galaxies, could suffer gas exhaustion by both gas depletion through star formation and some stripping process such as strangulation, \textit{also} in their ways \textit{towards} clusters.  Both mechanisms would lead to chemical enhancement (product of the increased star formation as compared to isolated dwarf galaxies in the first case, and of the lulled infall in the second). This scenario could explain the trends observed for the oxygen abundance and N/O ratio even in A1367 (a cluster of $M\sim10^{14}M_\sun$), as well as the elevated metallicities observed for some galaxies at $R>R_{200}$ in both Coma and A1367.

\citet{Carter2002} have invoked pressure confinement by the intracluster medium to explain the cluster-centric radial  gradient of the stellar metallicity for a sample of passive galaxies in Coma. \citet{Smith2009}, however, have argued that the observed gradient should be interpreted as a trend in age rather than metallicity. Younger passively evolving dwarfs could have arisen by the transformation of infalling field late-type galaxies, indicating that the build-up of the passive population is an ongoing process in the outskirts of the cluster, related to environment-driven processes \citep[regarding the infalling galaxy population in Coma, see also][]{Aguerri2004}.

In a subsequent work, \citet{Smith2011} have found that the cluster-centric age gradient for the red-sequence dwarfs in Coma is a global trend, and is not driven by the ongoing merger of the NGC 4839 group to the south west of Coma. These authors have compared their results with the predictions from simulated cluster assembly histories  \citep[the Millenium Simulation,][]{Springel2005} and have argued that in order to reproduce the strength of the age gradient observed for the red-sequence dwarfs in Coma, either a dominant burst or a gradual decline in the star-formation rate has to be invoked; models with very abrupt quenching would lead to shallower age trend. These findings support the scenario proposed here: recent star-formation, within a massive cluster like Coma, could yield the chemical enhancement of dwarf cluster  galaxies.

Finally, a mechanism for the chemical enrichment of cluster galaxies to be considered could be the presence of enriched inflows.
Historically it has been thought that there was a remarkable uniformity in the metal abundance of the ICM ($\sim 0.5 Z_\sun$) as a function of cluster mass, ranging from cooling-core to noncooling-core clusters (see \citet{Werner2008} for a review on the chemical enrichment in the ICM,  and \citet{Schindler2008} for a review on the processes proposed to explain this enrichment). However, recently, the spatially resolved analysis of the chemical composition of the ICM  has revealed that this is not uniformly enriched in metals. ICM abundance gradients are common in clusters, showing a peak in the central region and a decline outwards \citep[e.g.][]{DeGrandi2004,Leccardi2008,Lovisari2011}. The central metallicity of the ICM can reach even over solar values, associated to the presence of the brightest cluster galaxy (RPS could also play a role in the metallicity enhancement in cluster centers). The question arises as to whether infall of enriched material of the ICM could affect the metallicity of cluster galaxies. However, detailed modeling would be needed, regarding the cooling of the ICM and the accretion mechanisms, which is beyond the scope of the present work.

\section{SUMMARY AND CONCLUSIONS} \label{SUM}

In this work we have studied the chemical history of low-mass SF galaxies in four clusters in the local Universe. The sample clusters belong to a semi-spheric shell of the local Universe ($\delta\gtrsim$ -25 deg and $0.02<z<0.03$) and span a mass range from $10^{13}$ to $10^{15} M_\sun$. The regions studied cover the clusters' core up to 3$R_{200}$. We have been searching for the potential imprints of the cluster environment on the galaxy chemical enrichment.

We have used the latest SDSS spectroscopic release DR8. SF galaxies have been selected on the basis of their SDSS emission line fluxes, and a limit in magnitud has been applied to select dwarf galaxies. Considering low-mass galaxies, apperture biases are not expected to be important. We note that DR8 spectroscopic data have been corrected for the underlying stellar continuum;  this being an important improvement when studing nebular gas properties.  We have found that our SF dwarf galaxies show typical line ratios of normal \HII galaxies. 

Gas-phase metallicities of the O/H and N/O ratio, have been derived carefully using different empirical and model calibrations. The accurate mass estimates provided by SDSS DR8 have been used, to derive the MZR of the cluster galaxies. Well defined sequences have been found in the MZ plane, and we have observed a decrease of the scatter when the correction for the SFR has been applied. The value derived for the slope of the MZR is in agreement with the predictions of hydrodynamic models, which use momentum-driven winds to reproduce the MZR. Well defined sequences have also been derived in the N/O versus mass plane.

For the more massive clusters of this sample, Coma and A1367, the galaxies located at cluster-centric distances $R\le R_{200}$, are preferentially located at the upper part of the global sequences of O/H and  N/O versus mass. This increase in metallicity is mass dependent, being  higher at the lower mass bins, and in the core of Coma reaches on average $\sim0.15$ dex in O/H. This effect yields the flattening of the MZR for SF dwarf galaxies within the core of these massive clusters.

The metal enhancement of SF dwarfs in the cluster core has been found to be more important when considering the $R_{200}$ region of the most massive cluster Coma ($M\simeq10^{15} M_\sun$). Despite the general good relation of local galaxy density with cluster-centric distance, this effect appears diluted in terms of local galaxy density, suggesting that the relevant parameter able to affect the chemical evolution of SF dwarf galaxies should be the presence of a dense ICM.  

Finally, we have related the metallicity of our SF dwarf galaxies, with their HI mass content, derived using available 21 cm data. We then have compared with the predictions of the so-called ``closed-box`` model and the normal HI content of isolated couterparts, and we have found that SF dwarf galaxies in the cores of A1367 and Coma should be suffering an important ram-pressure stripping. 

We discuss that the properties of the ICM could be a key parameter to the chemical evolution of low-mass cluster galaxies. Efficient gas stripping (ram pressure or/and strangulation) and effective metal retention, due to the suppression of galactic winds, could lead to a different chemical enrichment scenario for cluster galaxies.

The present sample consists of four clusters, and although the trends observed could be insightful, meaningful coclusions could be taken out only by improving considerably statistics. In a future work we will investigate further the connection of the chemical enrichment of cluster galaxies with the properties of the ICM, for a larger set of clusters,  sampling a wide range of X-ray luminosities.

\acknowledgments

We thank Dr. O. Dors and Dr. E. Perez-Montero for sharing their models to derive galaxy metallicity. We also thank the anonymous referee for
useful suggestions that helped improving the paper. V.P. acknowledge financial support from the Spanish Ministerio de Ciencia e Innovaci\'on under grant FPU AP2006-04622. We also acknowledge financial support by the Spanish PNAYA project ESTALLIDOS (AYA2010-21887-C04-01) and CSD2006 00070 “1st Science with GTC” from the CONSOLIDER 2010 programme of the Spanish MICINN. 

Funding for SDSS-III has been provided by the Alfred P. Sloan Foundation, the Participating Institutions, the National Science Foundation, and the U.S. Department of Energy Office of Science. The SDSS-III web site is http://www.sdss3.org/. SDSS-III is managed by the Astrophysical Research Consortium for the Participating Institutions of the SDSS-III Collaboration including the University of Arizona, the Brazilian Participation Group, Brookhaven National Laboratory, University of Cambridge, University of Florida, the French Participation Group, the German Participation Group, the Instituto de Astrofisica de Canarias, the Michigan State/Notre Dame/JINA Participation Group, Johns Hopkins University, Lawrence Berkeley National Laboratory, Max Planck Institute for Astrophysics, New Mexico State University, New York University, Ohio State University, Pennsylvania State University, University of Portsmouth, Princeton University, the Spanish Participation Group, University of Tokyo, University of Utah, Vanderbilt University, University of Virginia, University of Washington, and Yale University. This research has made use of the NASA/IPAC Extragalactic Database (NED) which is operated by the Jet Propulsion Laboratory, California Institute of Technology, under contract with the National Aeronautics and Space Administration.

\include{biblio}
\end{document}

%% file: tbl1.tex
\begin{deluxetable}{lccccccccc}
\tabletypesize{\scriptsize}
\tablecaption{Cluster Properties\label{tbl1}}
\tablewidth{0pt}
\tablehead{
\colhead{Cluster} &\colhead{RA} &\colhead{DEC} &\colhead{z} &\colhead{$\sigma_\mathrm{v}$} &\colhead{m-M} &\colhead{Scale} &\colhead{$R_{200}$} &\colhead{$M_{cl}$ } &\colhead{$L_X$}\\
\colhead{} &\colhead{J2000} &\colhead{J2000} &\colhead{} &\colhead{(km s$^{-1}$)} &\colhead{(mag)} &\colhead{(Mpc deg$^{-1}$)} &\colhead{(Mpc)} &\colhead{($M_\sun$)} &\colhead{($10^{43}$ erg s$^{-1}$)}\\
\colhead{(1)} &\colhead{(2)} &\colhead{(3)} &\colhead{(4)} &\colhead{(5)} &\colhead{(6)} &\colhead{(7)} &\colhead{(8)} &\colhead{(9)} &\colhead{(10)}}
\startdata
A1656 & 12 59 48.7  & 27 58 50  & $0.0231$ 	& 1008 & 35.06 & 1.79 & 1.73 & 	$1.2\times 10^{15}$  &	$9.30\pm0.14$ 	\\	
A1367 & 11 44 29.5  & 19 50 21  & $0.022\phn$  	& 879  & 34.95 & 1.71 & 1.51 & 	$8.1\times 10^{14}$  &	$2.30\pm0.08$   \\	
A779  & 09 19 50.8  & 33 46 17  & $0.0225$ 	& 339  & 34.92 & 1.68 & 0.58 & 	$4.6\times 10^{13}$  &	$0.29\pm0.04$ 	\\	
A634  & 08 14 33.7  & 58 02 52  & $0.0265$ 	& 391  & 35.23 & 1.94 & 0.67 & 	$7.1\times 10^{13}$  &	$<0.08$ 	\\	

\enddata
\tablecomments{Column 1: Cluster; Column 2: Right ascension in hours, minutes, and seconds of cluster center as given in NED; Column 3: declination, in degrees, arcminutes, and arcseconds of cluster center (NED); Column 4: cluster mean redshift (NED); Column 5: cluster velocity dispersion in km s$^{-1}$ \citep{Struble1999}; 
Column 6: cluster distance modulus in magnitudes (NED); Column 7: cluster scale in Mpc deg$^{-1}$; Column 8: cluster $R_{200}$ in Mpc as derived using Equation (8) of \citet{Finn2005}; Column 9: cluster mass in $M_\sun$, as derived using Equation (10) of \citet{Finn2005}; Column 10:  ROSAT X-ray luminosity $L_X$ in units $10^{43}$ erg s$^{-1}$  \citep{Ledlow2003}.}
\end{deluxetable}

%% file: tbl2.tex
\begin{deluxetable}{lcccccccc}
\tabletypesize{\scriptsize}
\tablecaption{cluster regions considered\label{tbl2}}
\tablewidth{0pt}
\tablehead{
\colhead{Cluster}  &\colhead{Area} &\colhead{$z$} &\colhead{Total} &\colhead{Dwarf} &\colhead{SF} &\colhead{Dwarf SF} &\colhead{Dwarf} &\colhead{Dwarf SF}\\
\colhead{}  &\colhead{deg$^2$} &\colhead{} &\colhead{} &\colhead{} &\colhead{} &\colhead{} &\colhead{$R_{200}$} &\colhead{$R_{200}$}}
\startdata
A1656  & $5.7\times5.7$   & 0.015-0.0323  & 1017 	& 616 & 194  & 148 & 293 & 25\\
A1367  & $5.2\times5.2$   & 0.015-0.0323  & 564  	& 356 & 238  & 191 & 107 & 41\\
A779   & $2\times2$       & 0.0189-0.0275 & 106 	& 66  & 32   & 31  & 26  & 7\\
A634   & $2\times2$       & 0.0246-0.0293 & 97  	& 60  & 34   & 26  & 24  & 10 \\ 
\enddata
\tablecomments{Column 1: Cluster; Column 2: Square area studied in deg$^2$; Column 3: The redshift range considered for each cluster; Column 4: Total number of galaxies in the considered area and velocity range; Column 5: Number of dwarf ($zmag>15$) galaxies; Column 6: Number of SF galaxies; Column 7: Number of SF dwarf galaxies; Column 8: Number of dwarf galaxies at distances $R\le R_{200}$ from the cluster center; Column 9: Number of SF dwarf galaxies at distances $R\le R_{200}$ from the cluster center.}
\end{deluxetable}

%% file: fig1.tex
\begin{figure}
\center
\includegraphics[width=7.5cm]{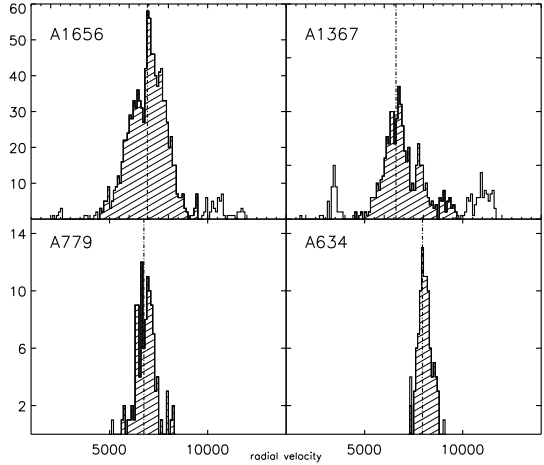}
\caption{The radial velocity histogram for all galaxies with SDSS-DR8 spectroscopic data within $3R_{200}$ from the center of each cluster. The velocity bin is 100 km s$^{-1}$. The dashed line represents the mean cluster radial velocity as given in NED (Coma: cz=6930 km s$^{-1}$; A1367: cz=6600 km s$^{-1}$; A779: 6750 km s$^{-1}$; A634: 7950 km s$^{-1}$). The dashed region indicates the adopted velocity range at $\pm3\sigma_\mathrm{v}$ around the mean cluster velocity. \label{fig1}}
\end{figure}

%

%% file: fig2.tex
\begin{figure}
\center
\includegraphics[width=7.5cm]{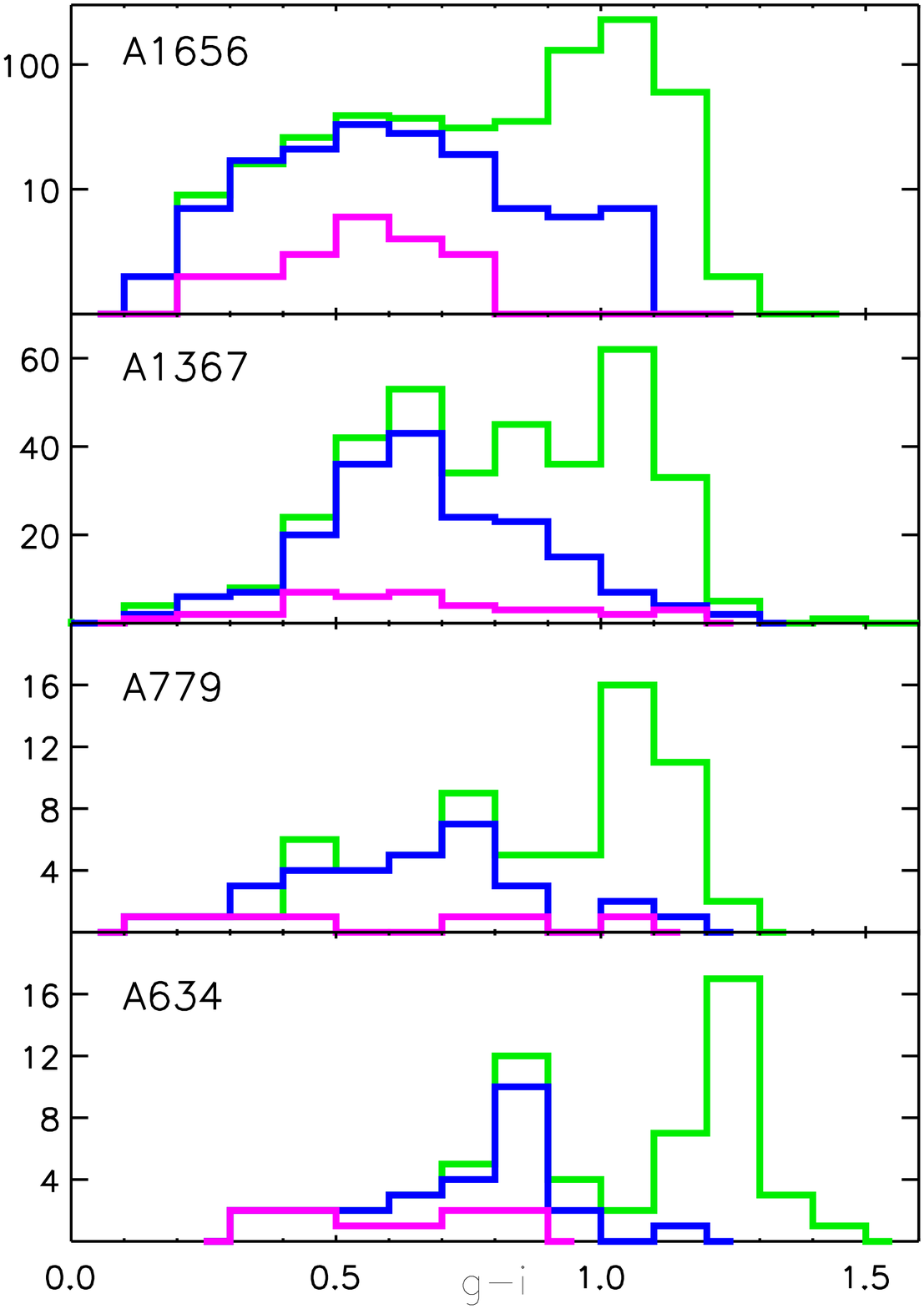}
\caption{The color g-i histogram of dwarf galaxies (green line), SF dwarf  galaxies (blue line), and SF dwarf  galaxies to $R\le R_{200}$ (magenta line) for our clusters sample.  Logarithmic scale is used for A1656. No  correction has been performed for the galactic extinction, and the shift observed for A634 towards redder colors  is consistent with the slightly higher galactic extinction suffered by this cluster. \label{fig2}}
\end{figure}

%

%% file: tbl3.tex
\begin{deluxetable}{lcccc}
\tabletypesize{\scriptsize}
\tablecaption{Galaxies with \OII3727 measures\label{tbl3}}
\tablewidth{0pt}
\tablehead{
\colhead{Cluster} &\colhead{DR8} &\colhead{New} &\colhead{Total} &\colhead{All}}
\startdata
A1656 & 35 & 52 & 87 & 149\\
A1367 & 42 & 48 & 90 & 194\\
\enddata
\end{deluxetable}

%% file: fig3.tex
\begin{figure}
\includegraphics[width=7.5cm]{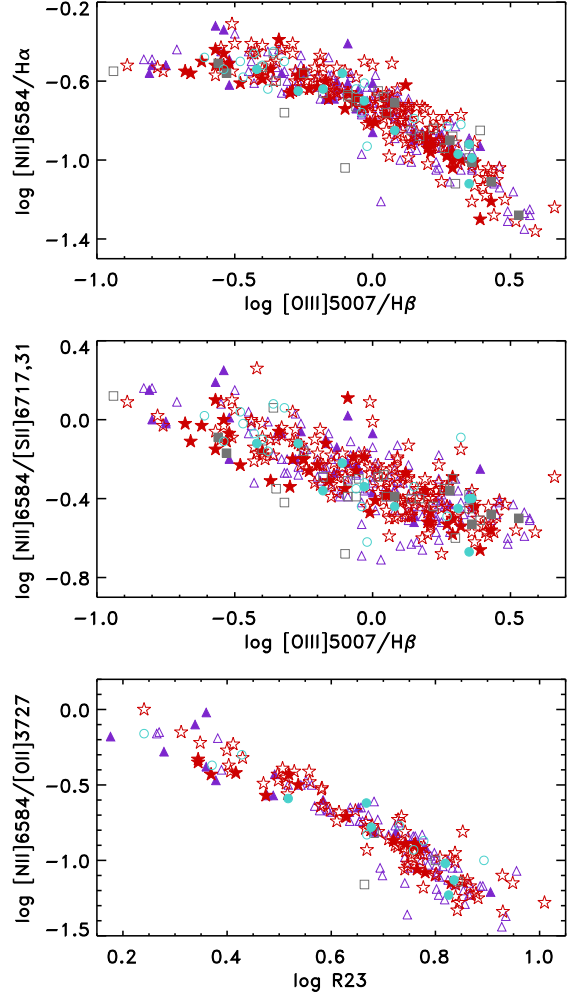}
\caption{Five line ratios combined into three generally used diagnostic diagrams for SF dwarf galaxies in Coma (triangles), A1367 (stars), A779 (squares), and A634 (circles). Filled symbols correspond to galaxies at $R\le R_{200}$.  Our sample galaxies show values typical of normal HII galaxies.\label{fig3}}
\end{figure}

%% file: fig4.tex
\begin{figure*}
\center
\includegraphics[width=7.5cm]{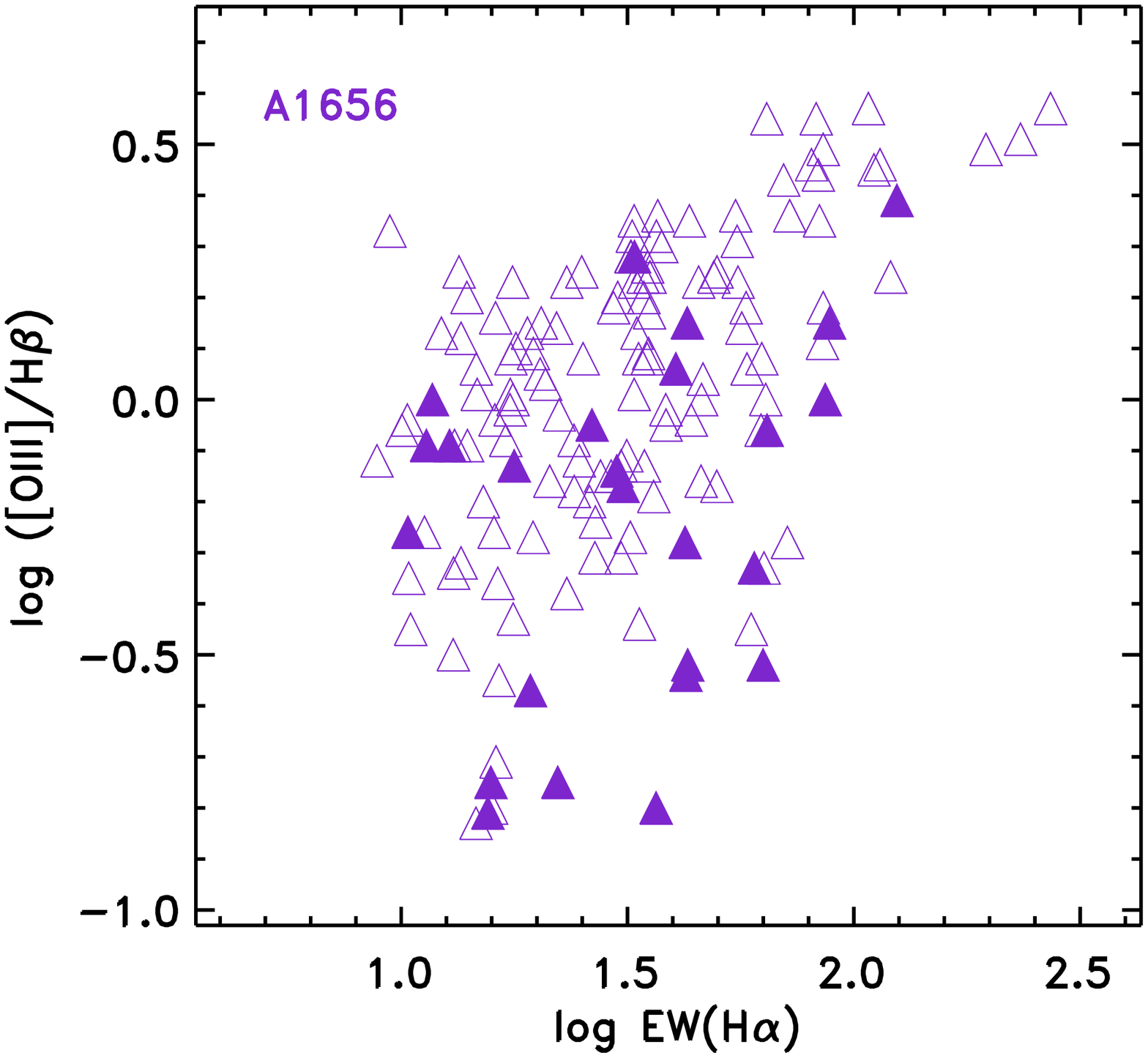}
\includegraphics[width=7.5cm]{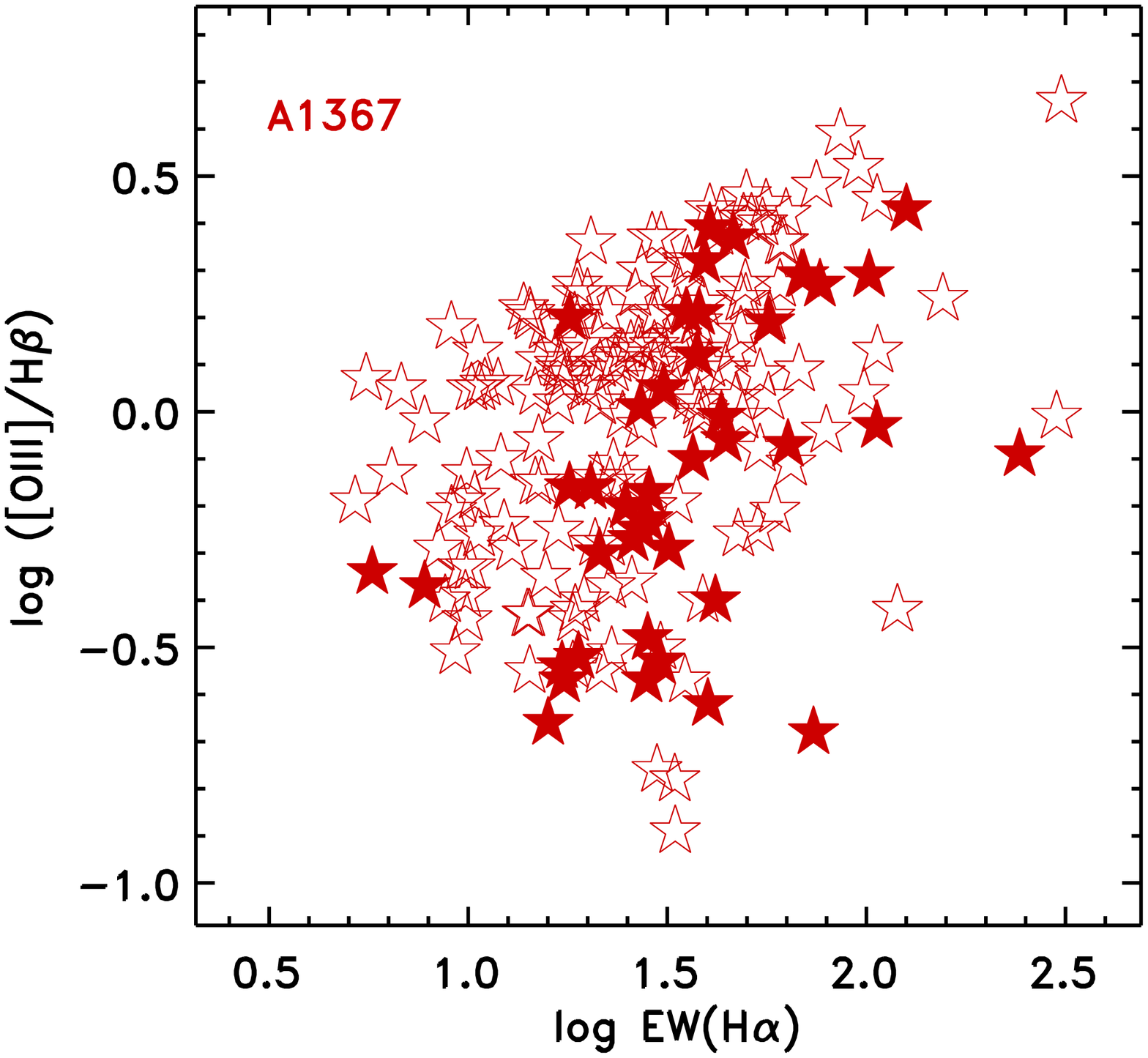}
\caption{The \OIII/H$\beta$ versus EW(H$\alpha$) for  SF dwarf galaxies in Coma (A1656, triangles) and A1367 (stars). Filled symbols represent galaxies at $R\le R_{200}$. \label{fig4}}
\end{figure*}

%% file: fig5.tex
\begin{figure*}
\center
\includegraphics[height=5.3cm]{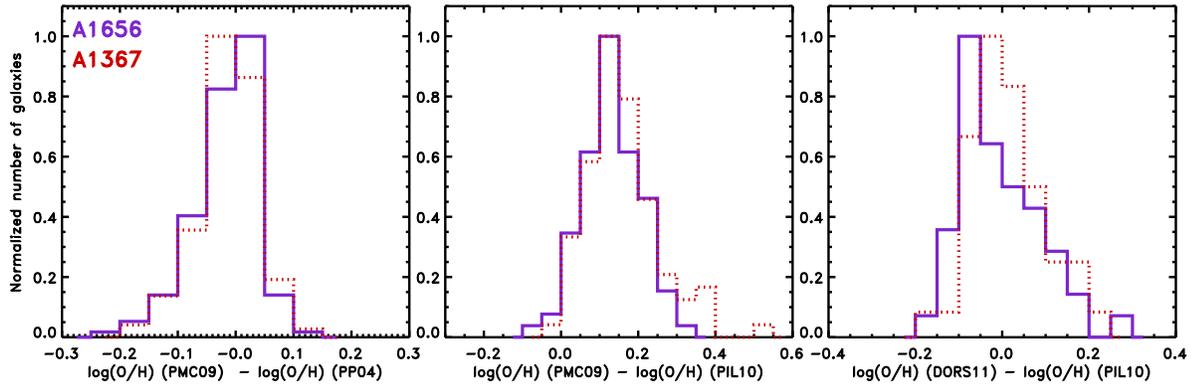}
\caption{Left: The difference of 12+log(O/H) derived using PMC09 and PP04 calibrations. A1656: continuous, A1367: dashed line. Middle: The same between PMC09 and PIL10 calibrations, for galaxies with \OII measured. Right: The same using \citet{Dors2011} models and PIL10 calibration (for galaxies with \OII measured and 12+log(O/H)$_\mathrm{DORS11}>$8.2).\label{fig5}}
\end{figure*}

%

%% file: fig6.tex
\begin{figure}
\center
\includegraphics[width=7.5cm]{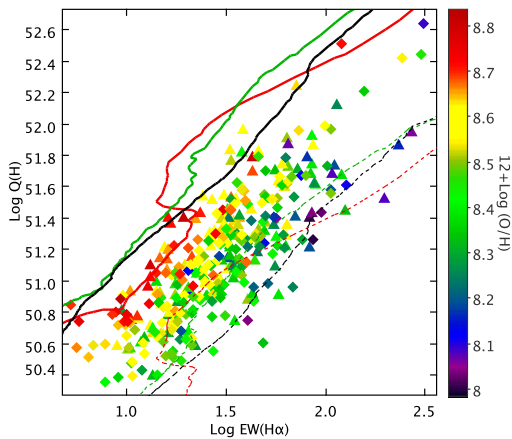}
\caption{The number of ionized photons Q(H), as measured by the \Ha flux of the SF region covered by the SDSS fibre, for our sample of SF dwarf galaxies in Coma (triangles) and A1367 (diamonds), versus the EW(H$\alpha$), color coded to the derived galaxy metallicity. Lines represent the Starburst99 models \citep{Leitherer1999} for instantaneous bursts of total mass in massive stars $M_\star=10^7 M_\sun$ (continuous) and $M_\star=10^6 M_\sun$ (dashed) and metallicities Z=0.02 (in red), Z=0.008 (green), and Z=0.004 (black). \label{fig6}}
\end{figure}

%% file: tbl4.tex
\begin{deluxetable}{lccc}
\tabletypesize{\scriptsize}
\tablecaption{Slope of MZR fit\label{tbl4}}
\tablewidth{0pt}
\tablehead{
\colhead{Cluster}  &\colhead{$R\leq R_{200}$} &\colhead{$R> R_{200}$} &\colhead{All}}
\startdata
A1656  & $0.19\pm0.03$  &  $0.33\pm0.03$  & $0.30\pm0.02$ \\
A1367  & $0.28\pm0.03$  &  $0.36\pm0.02$  & $0.33\pm0.02$\\
A779   & $0.30\pm0.11$  &  $0.28\pm0.05$  & $0.29\pm0.04$\\
A634   & $0.37\pm0.15$  &  $0.29\pm0.07$  & $0.32\pm0.05$\\
\enddata
\end{deluxetable}

%% file: fig7.tex
\begin{figure*}
\center
\includegraphics[width=8cm]{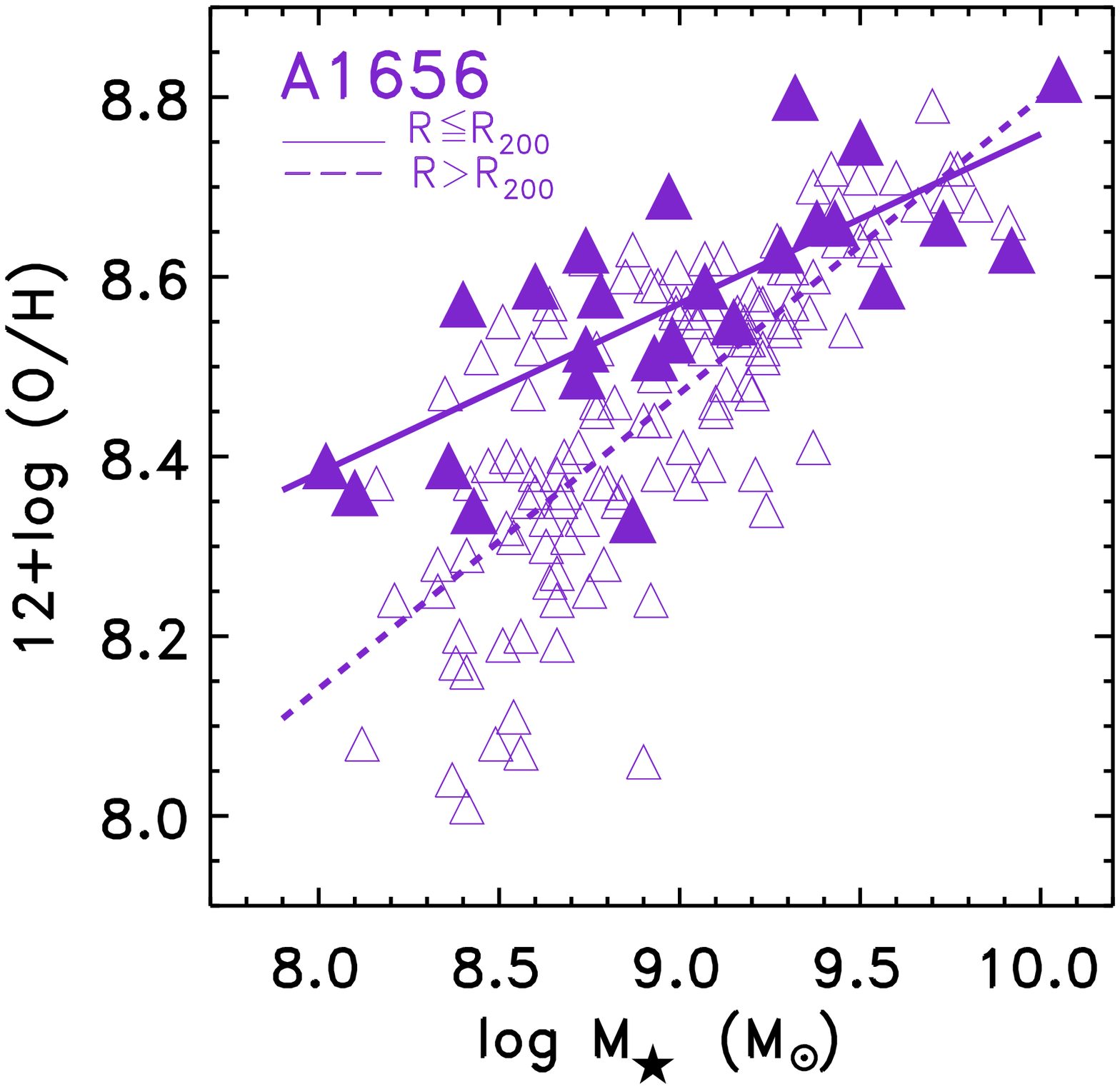}\vspace{-0.5cm}
\includegraphics[width=8cm]{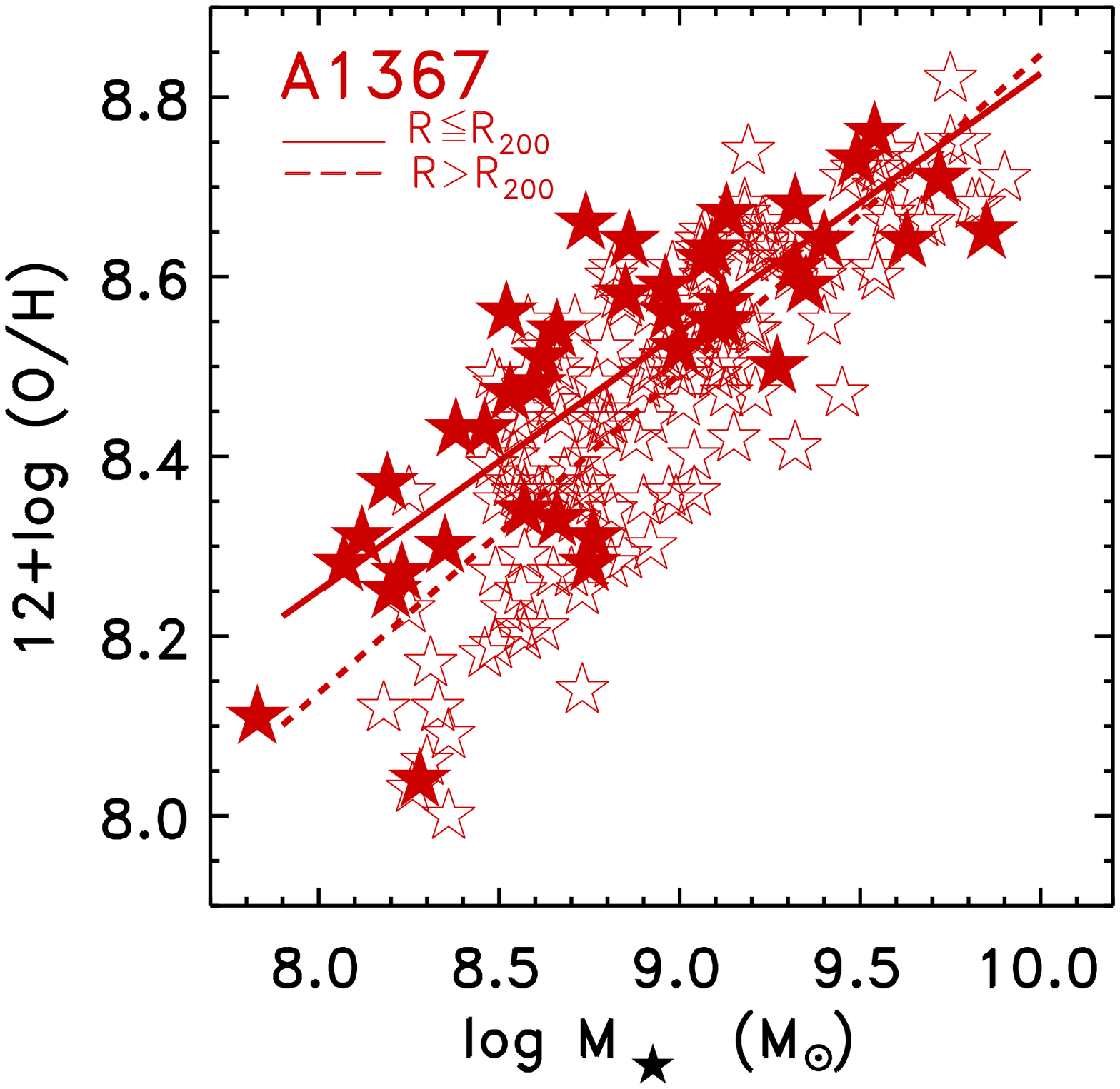}
\includegraphics[width=8cm]{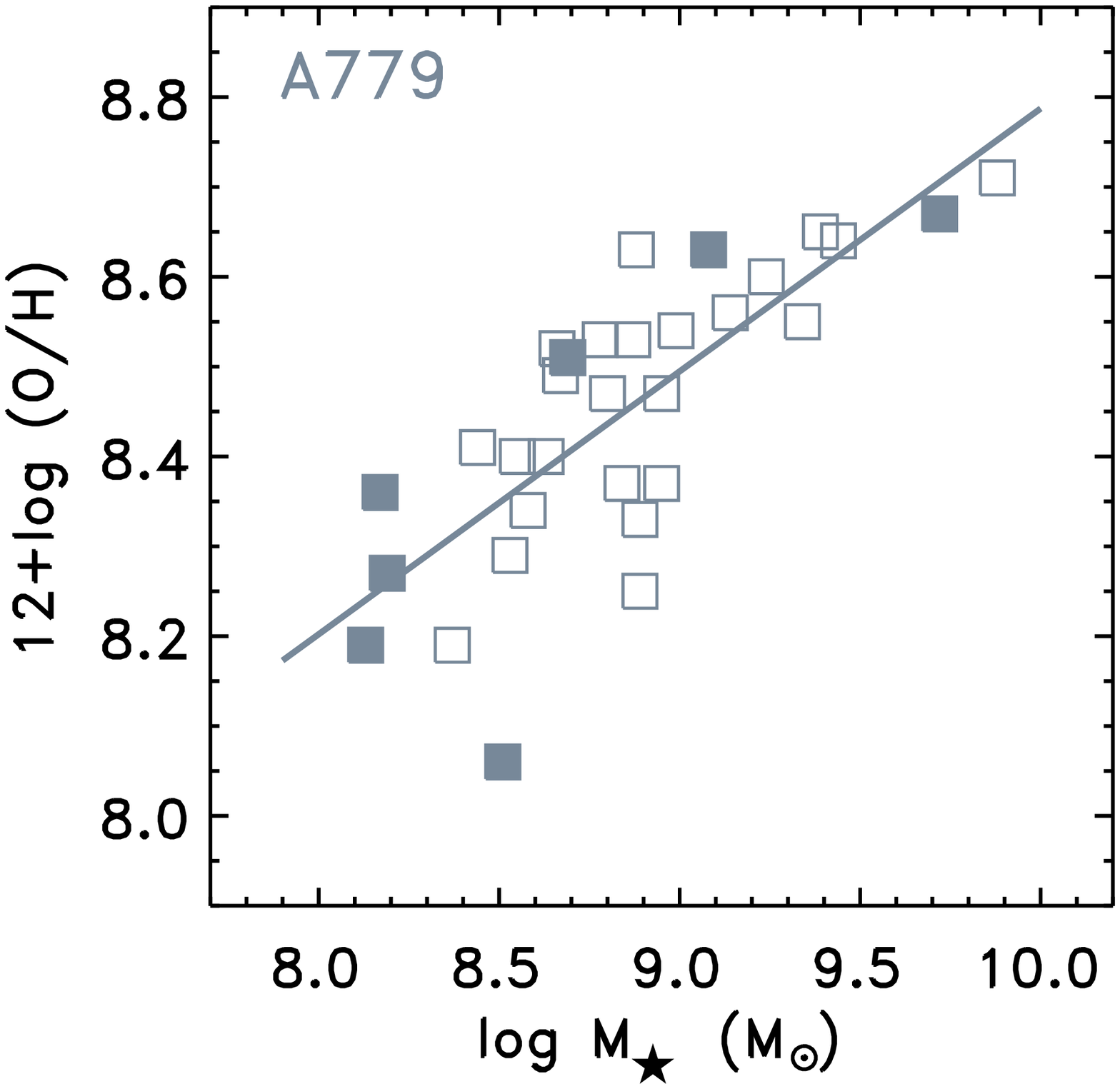}
\includegraphics[width=8cm]{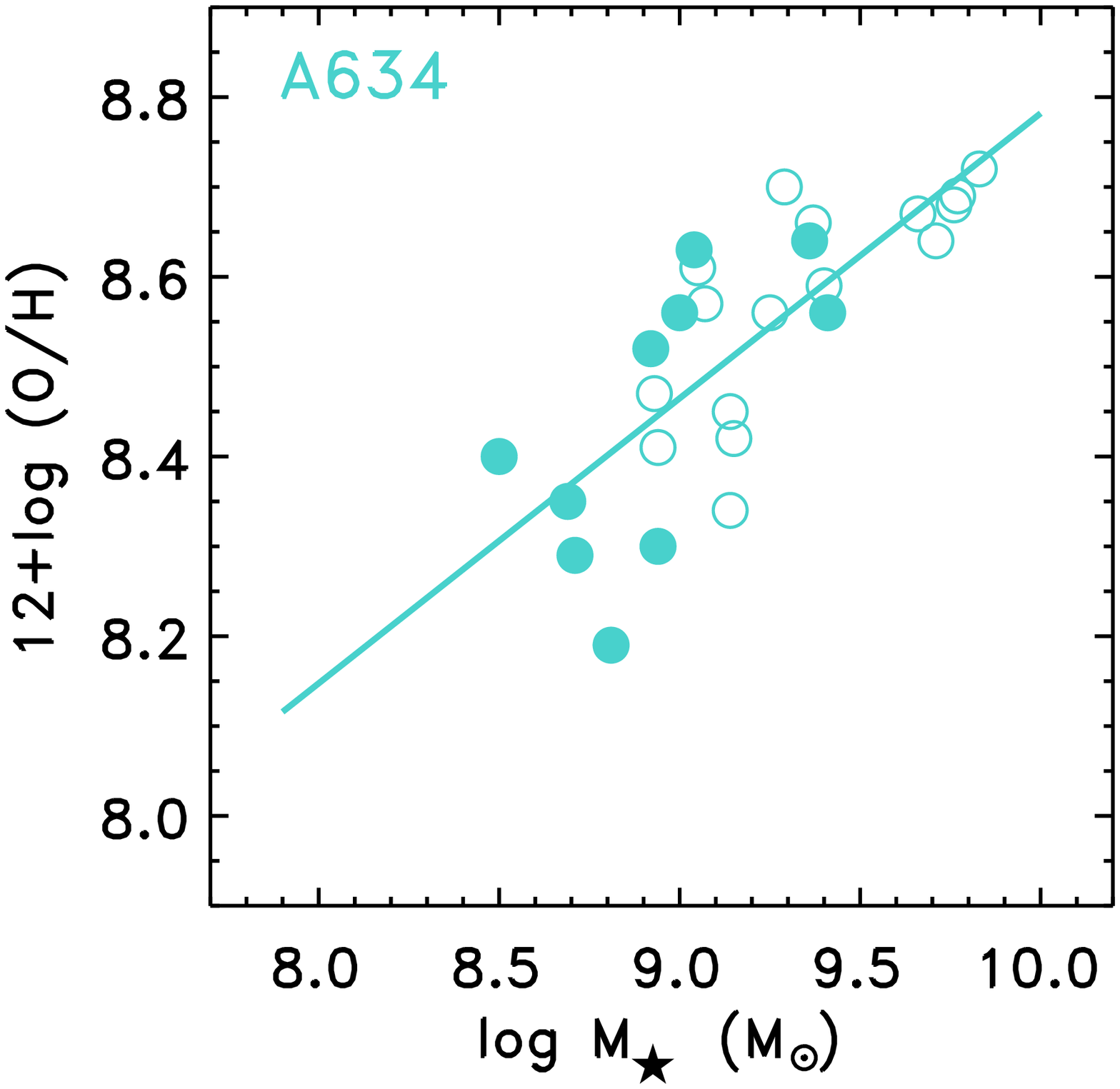}
\caption{The oxygen abundance 12+log(O/H) versus stellar mass for  SF dwarf galaxies in Coma (A1656, triangles), A1367 (stars), A779 (squares), and A634 (circles). Filled symbols correspond to galaxies at $R\le R_{200}$. For Coma and A1367: the continuous line is the linear fit for galaxies at $R\le R_{200}$ and the dashed line the fit for galaxies at $R>R_{200}$. For A779 and A634 the continuous line is the linear fit considering all galaxies. \label{fig7}}
\end{figure*}

%% file: fig8.tex
\begin{figure*}
\center
\includegraphics[width=8cm]{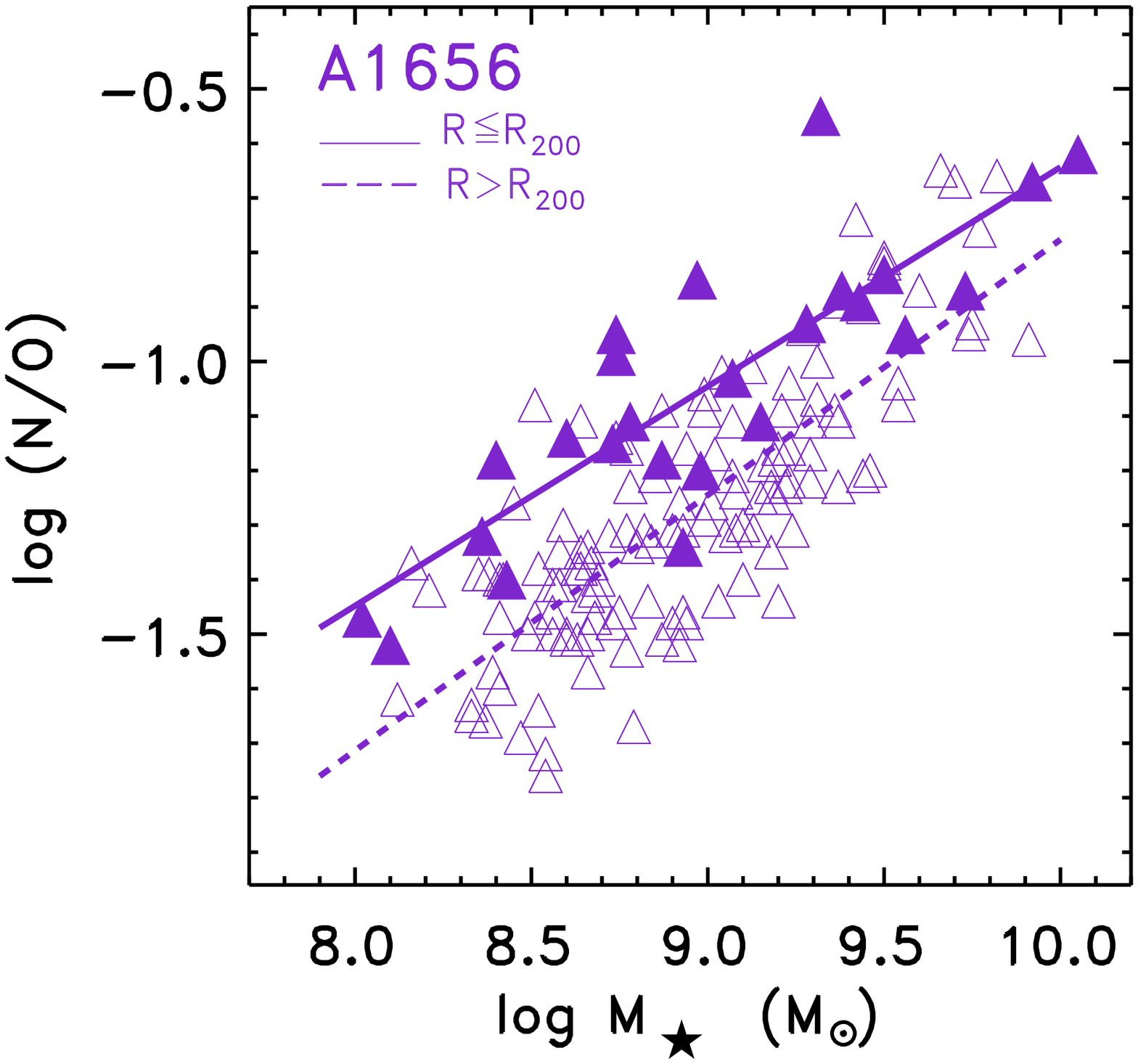}\vspace{-0.5cm}
\includegraphics[width=8cm]{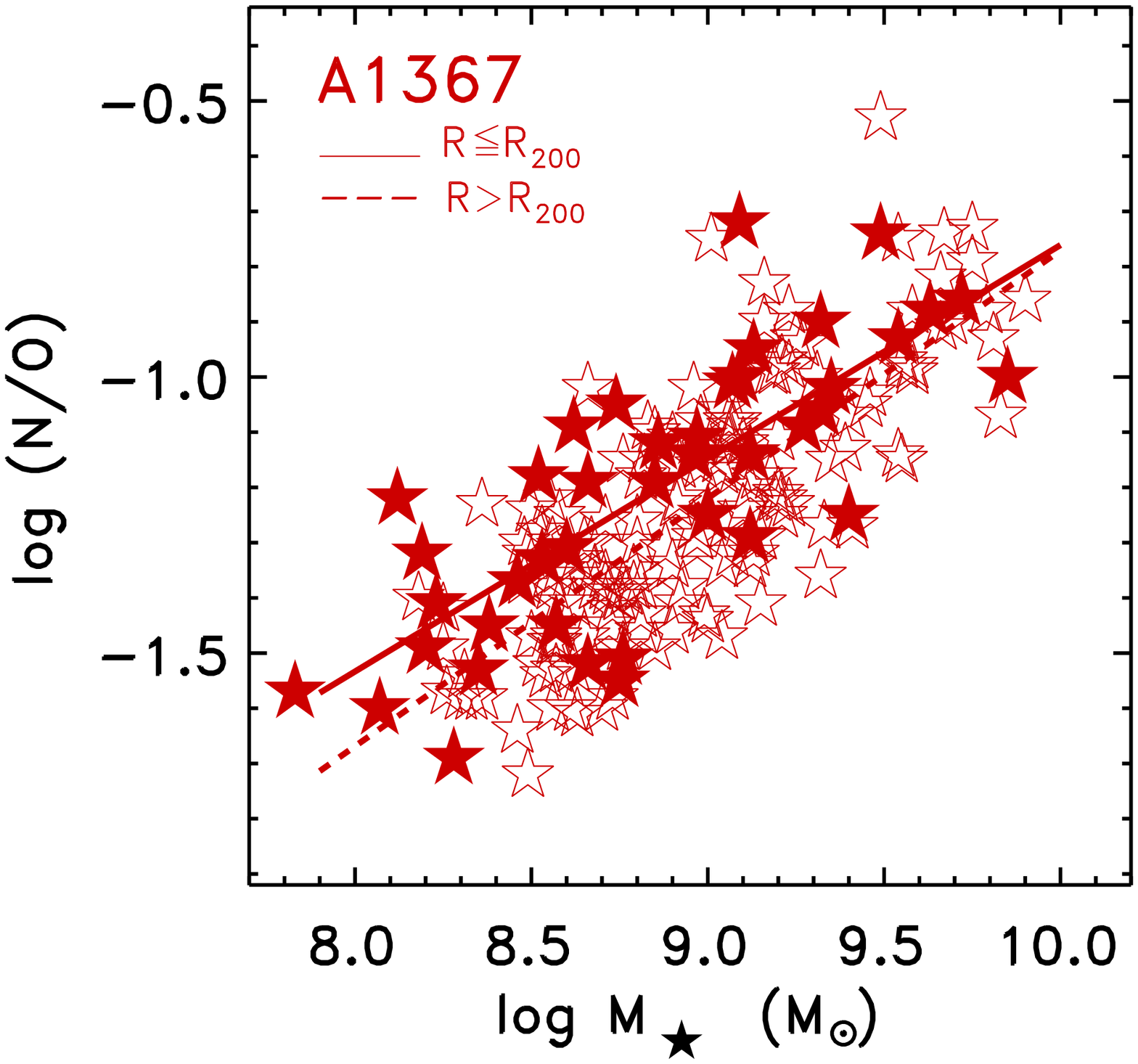}
\includegraphics[width=8cm]{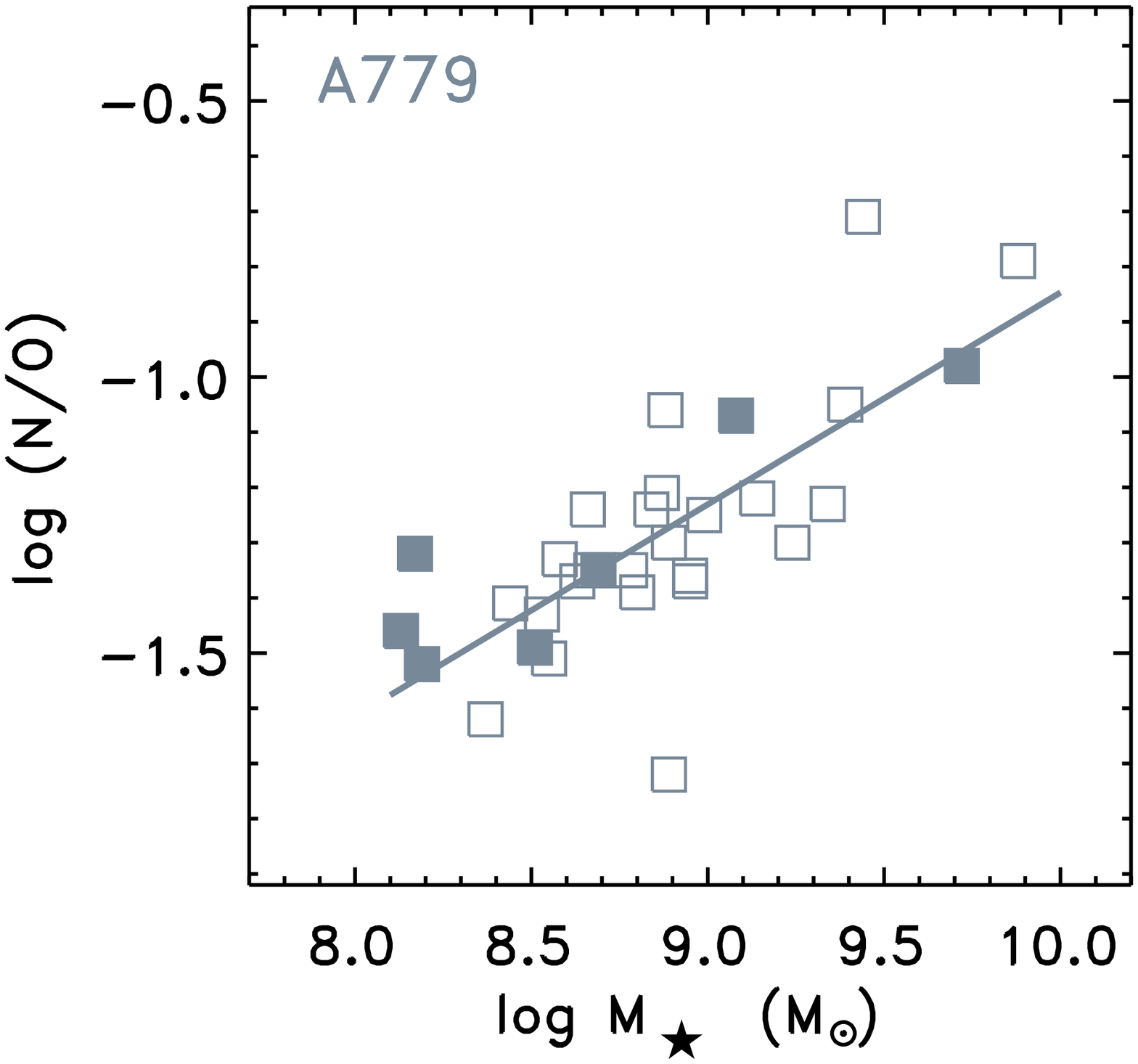}
\includegraphics[width=8cm]{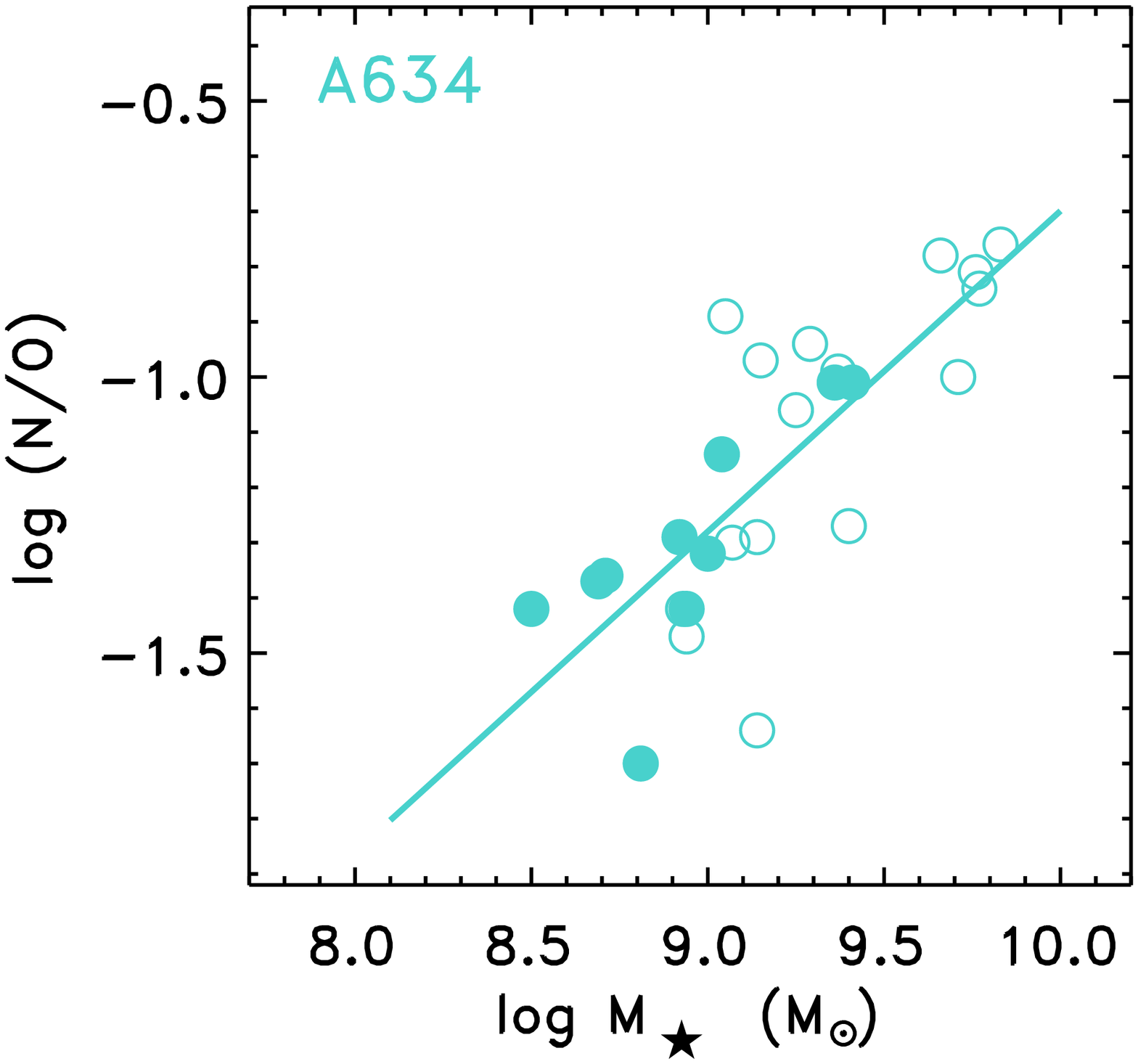}
\caption{N/O versus stellar mass for dwarf SF galaxies in Coma (triangles), A1367 (stars), A779 (squares), and A634 (circles). Filled symbols correspond to galaxies inside $R_{200}$.  For Coma and A1367: the continuous line is the linear fit for galaxies at $R\le R_{200}$ and the dashed line the fit for galaxies at $R>R_{200}$. For A779 and A634 the continuous line is the linear fit considering all galaxies. \label{fig8}}
\end{figure*}

%% file: fig9.tex
\begin{figure*}
\center
\includegraphics[width=8cm]{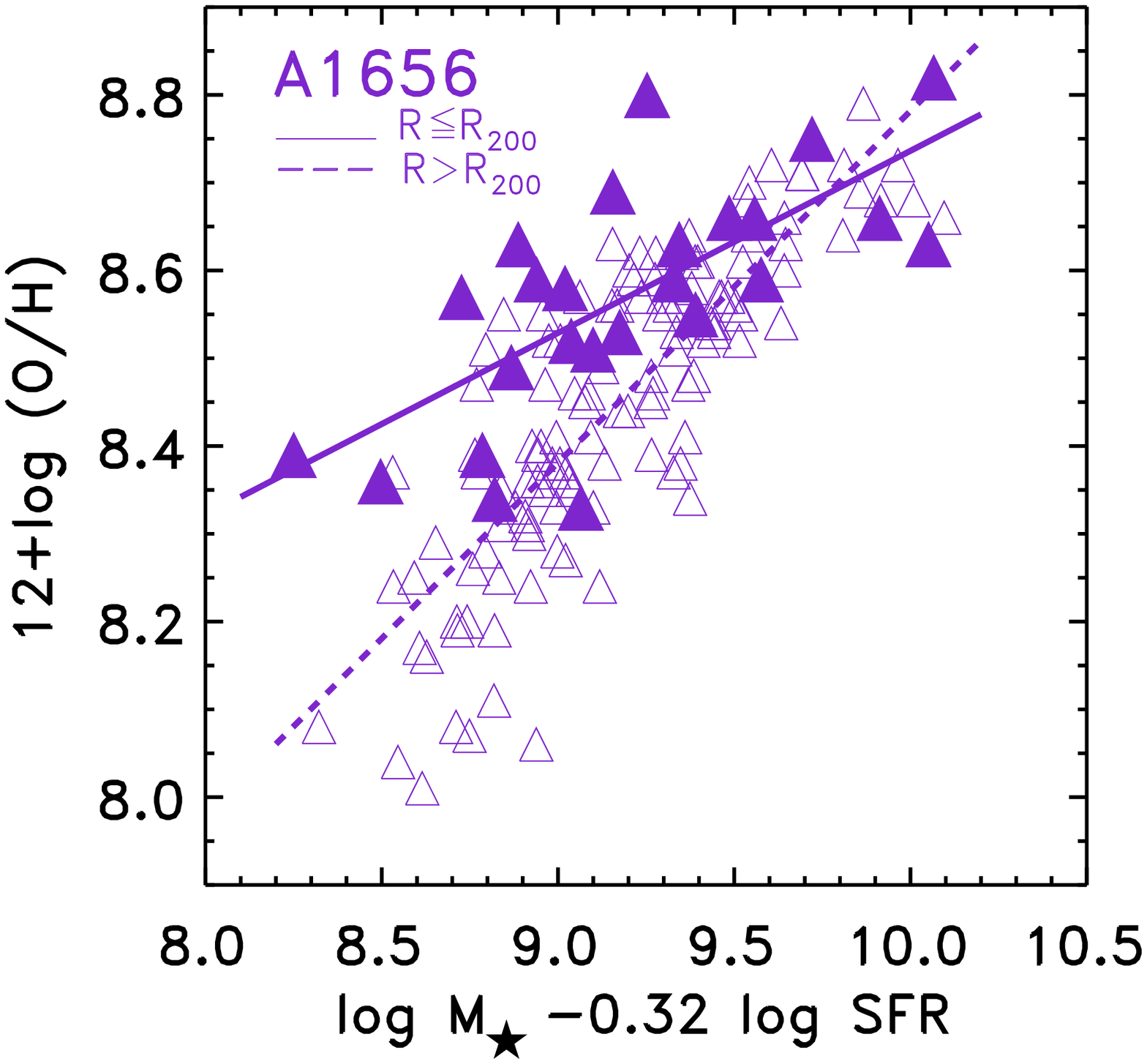}
\includegraphics[width=8cm]{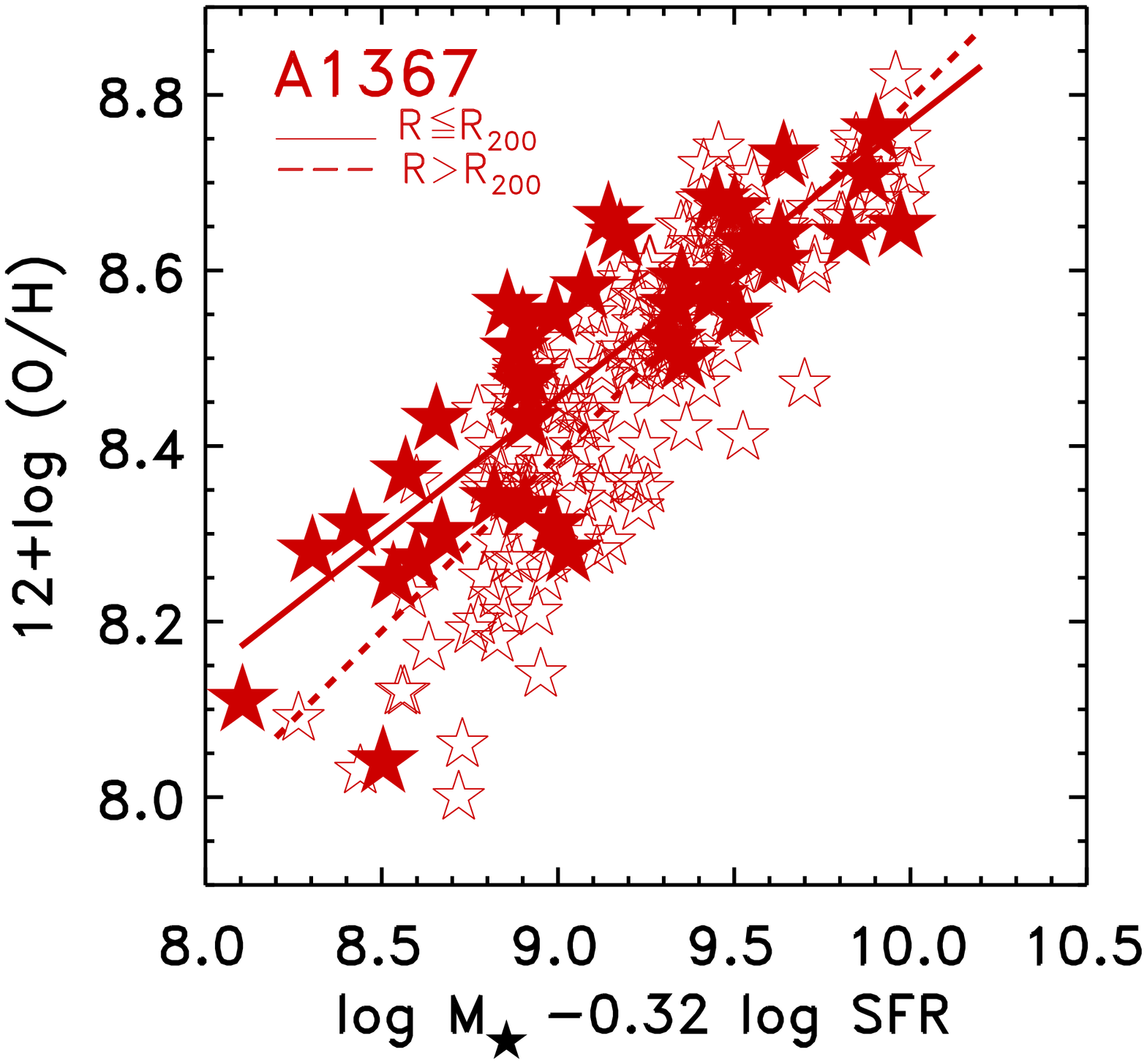}
\caption{The oxygen abundance 12+log(O/H) versus the quantity $\mu=\log M_\star-0.32\log$(SFR) for  SF dwarf galaxies in Coma (A1656, triangles) and A1367 (stars). Filled symbols correspond to galaxies at $R\le R_{200}$.  The continuous line is the linear fit for galaxies at $R\le R_{200}$ and the dashed line the fit for galaxies at $R>R_{200}$.\label{fig9}}
\end{figure*}

%% file: fig10.tex
\begin{figure}
\center
\includegraphics[width=8cm]{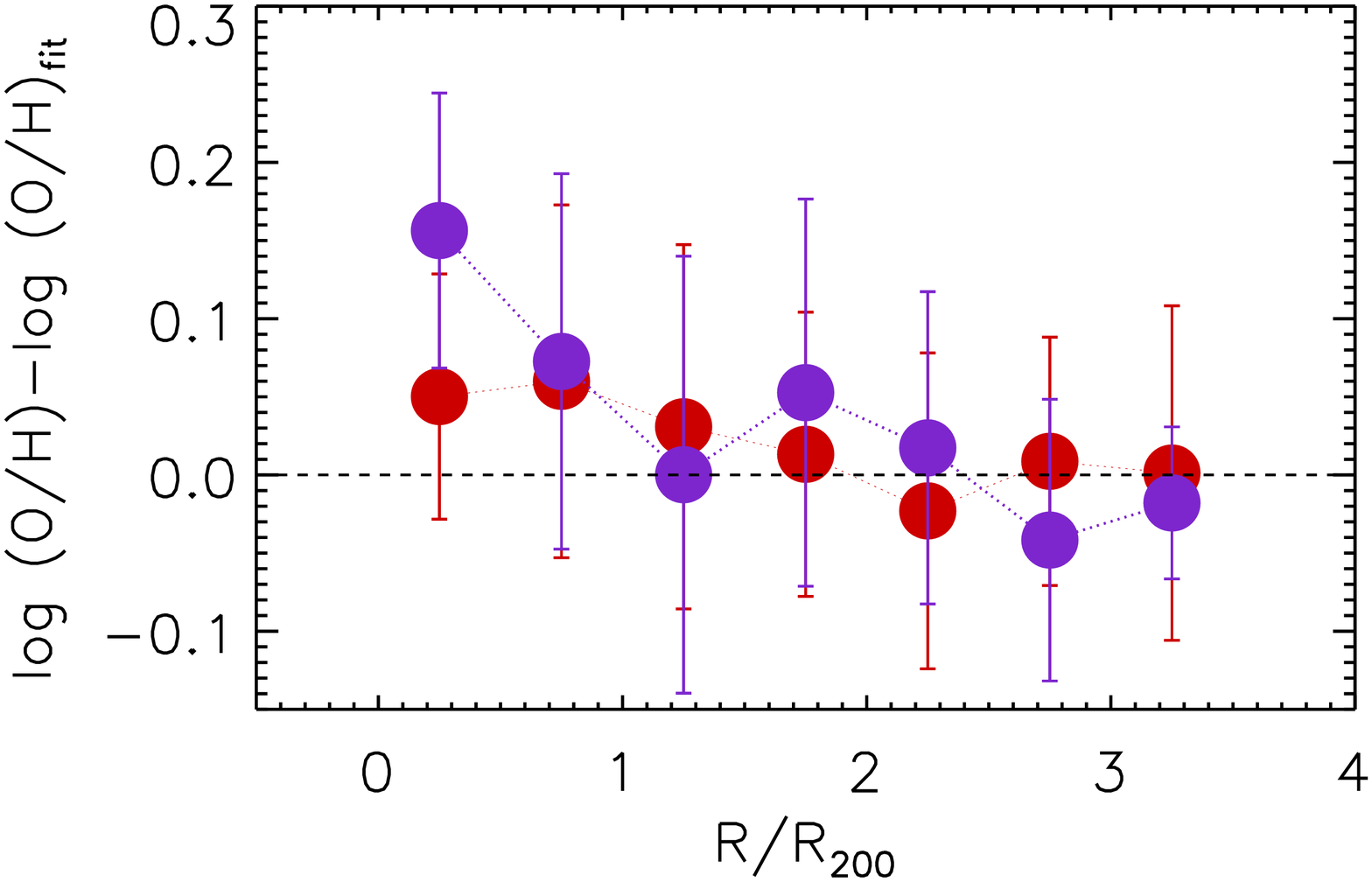}\vspace{-0.5cm}
\includegraphics[width=8cm]{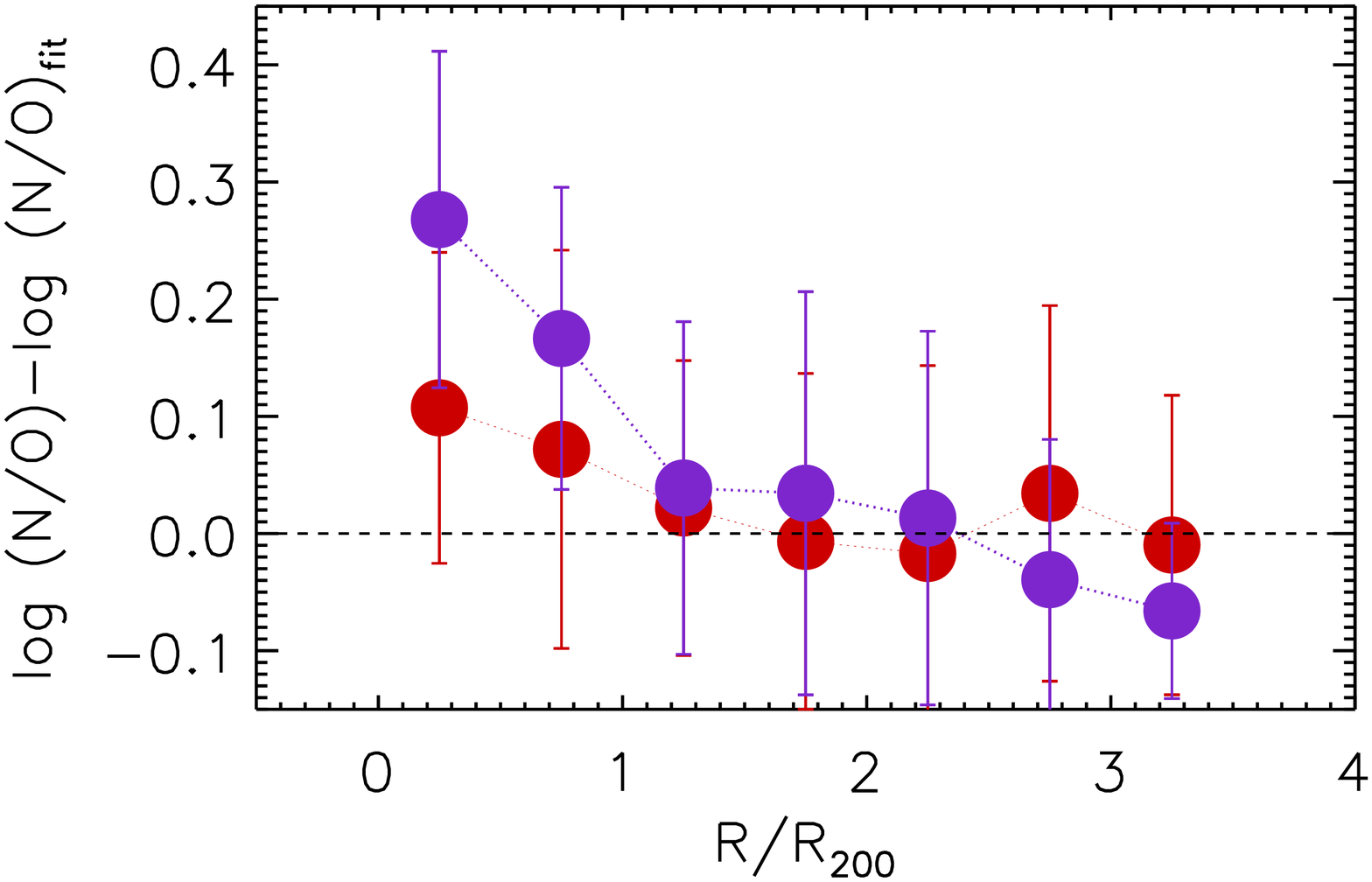}
\caption{Upper panel: The difference of the derived oxygen abundance 12+log(O/H), for each galaxy, with the oxygen abundance given by the bisector linear fit 12+log(O/H)$_\mathrm{fit}$, as a function of the cluster-centric radial distance $R/R_{200}$, in a bin of 0.5 $R_{200}$. Lower panel: The same for log(N/O). The blue points correspond to A1565 and the red poins to A1367. \label{fig10}}
\end{figure}

%% file: fig11.tex
\begin{figure}
\center
\includegraphics[width=8cm]{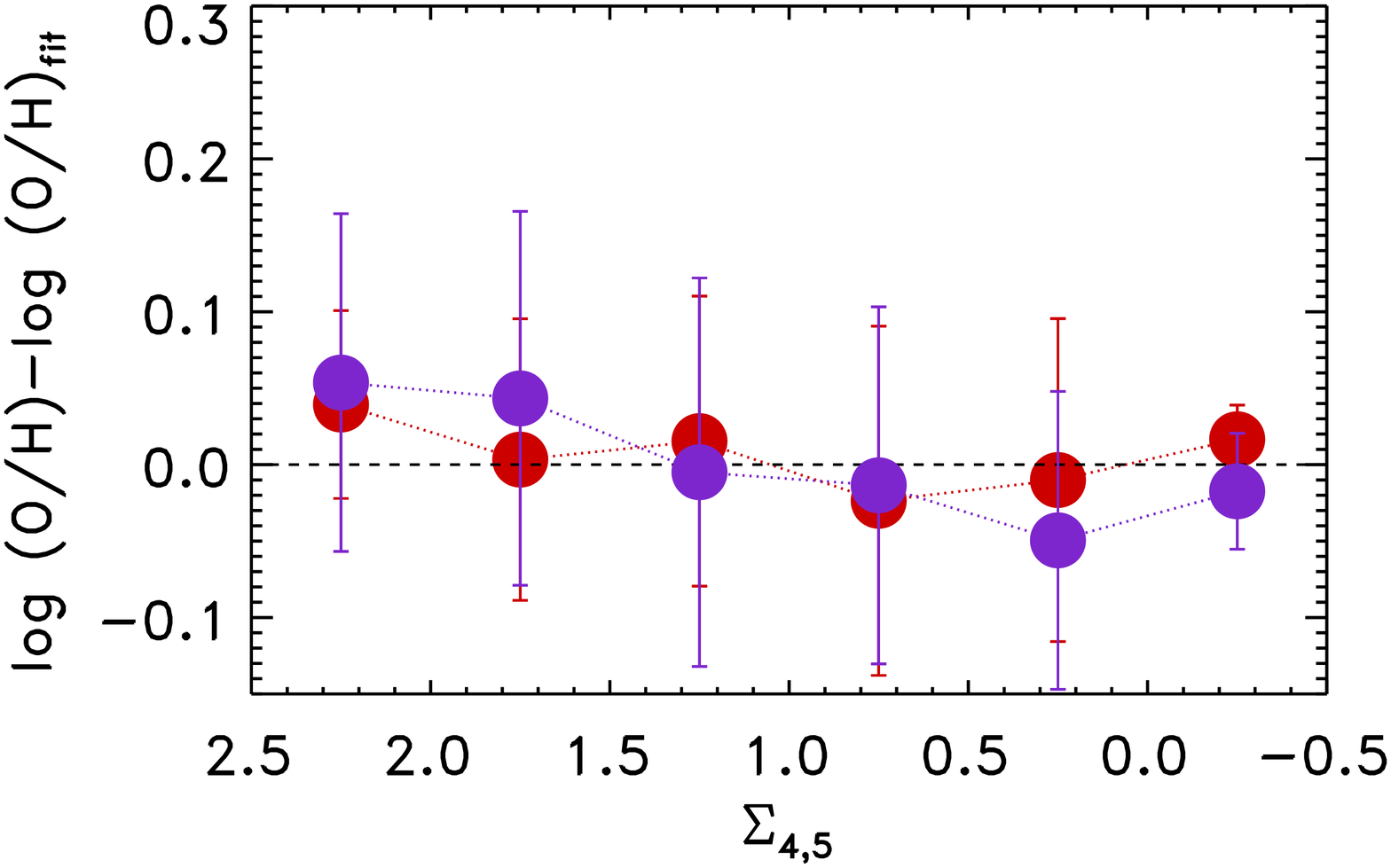}\vspace{-0.5cm}
\includegraphics[width=8cm]{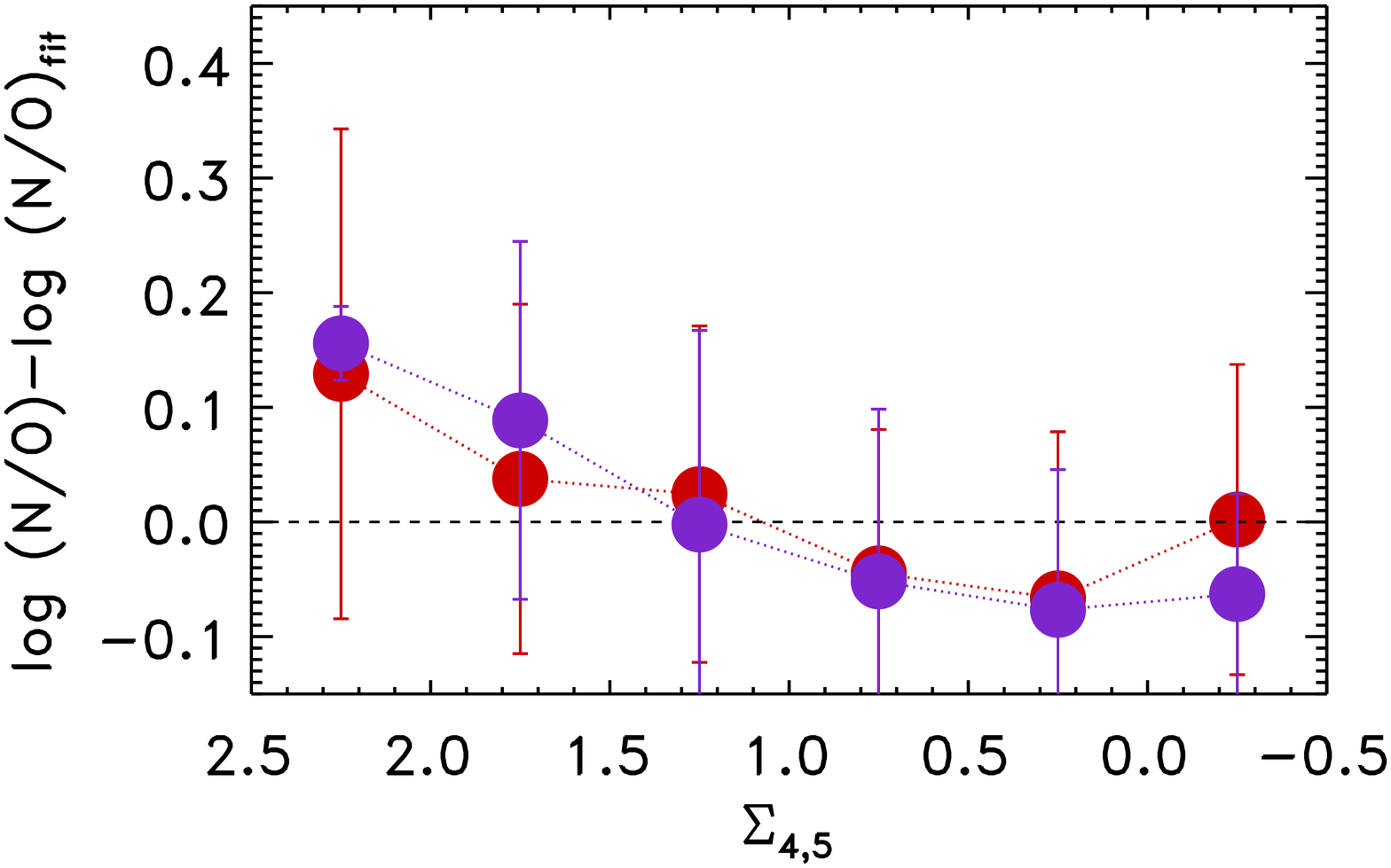}
\caption{The same as in Figure \ref{fig10}, as a function of the local galaxy density $\Sigma_{4,5}$ in a bin of 0.5 dex. \label{fig11}}
\end{figure}

%% file: fig12.tex
\begin{figure*}
\center
\includegraphics[width=7.5cm]{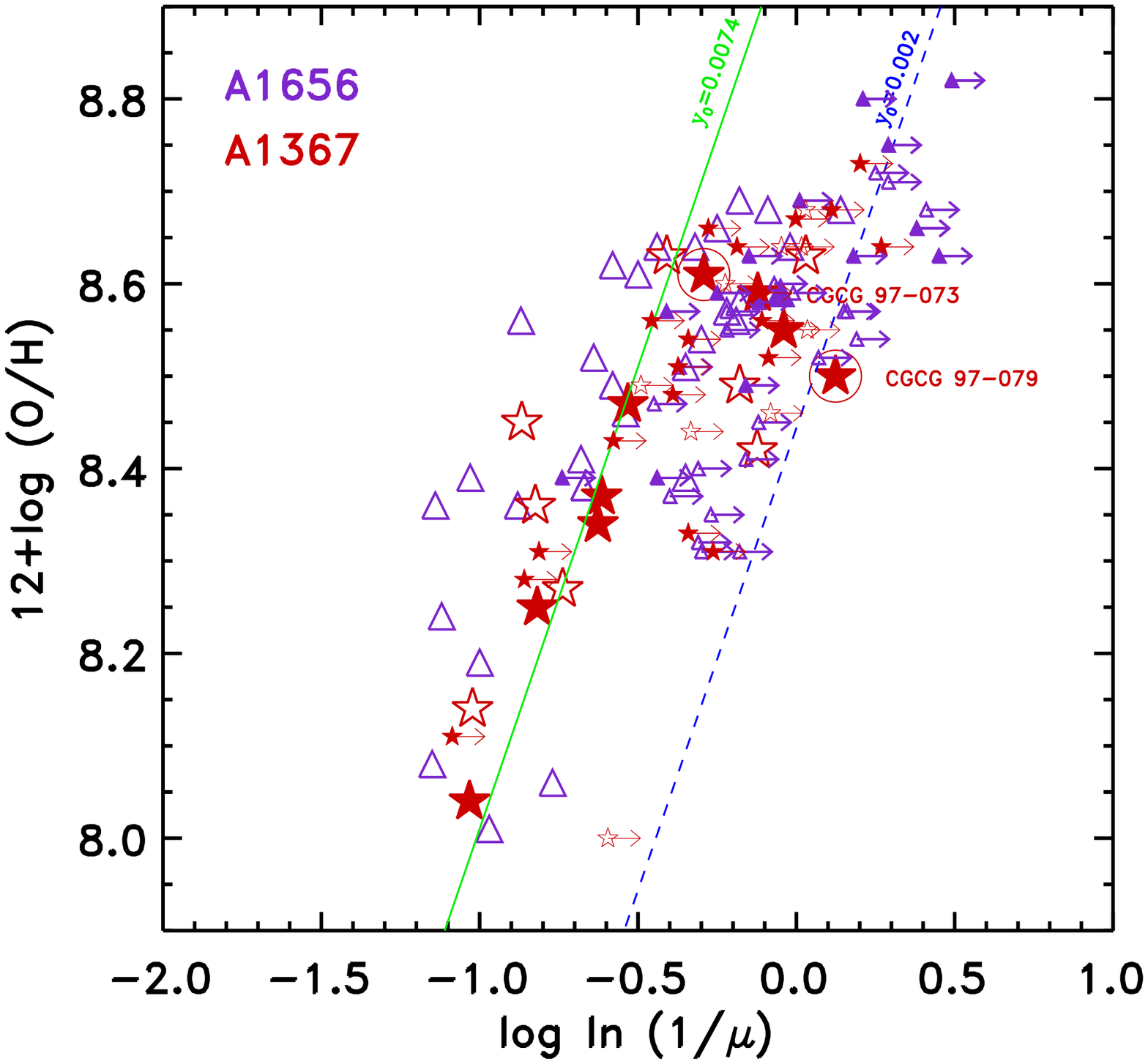}
\includegraphics[width=7.5cm]{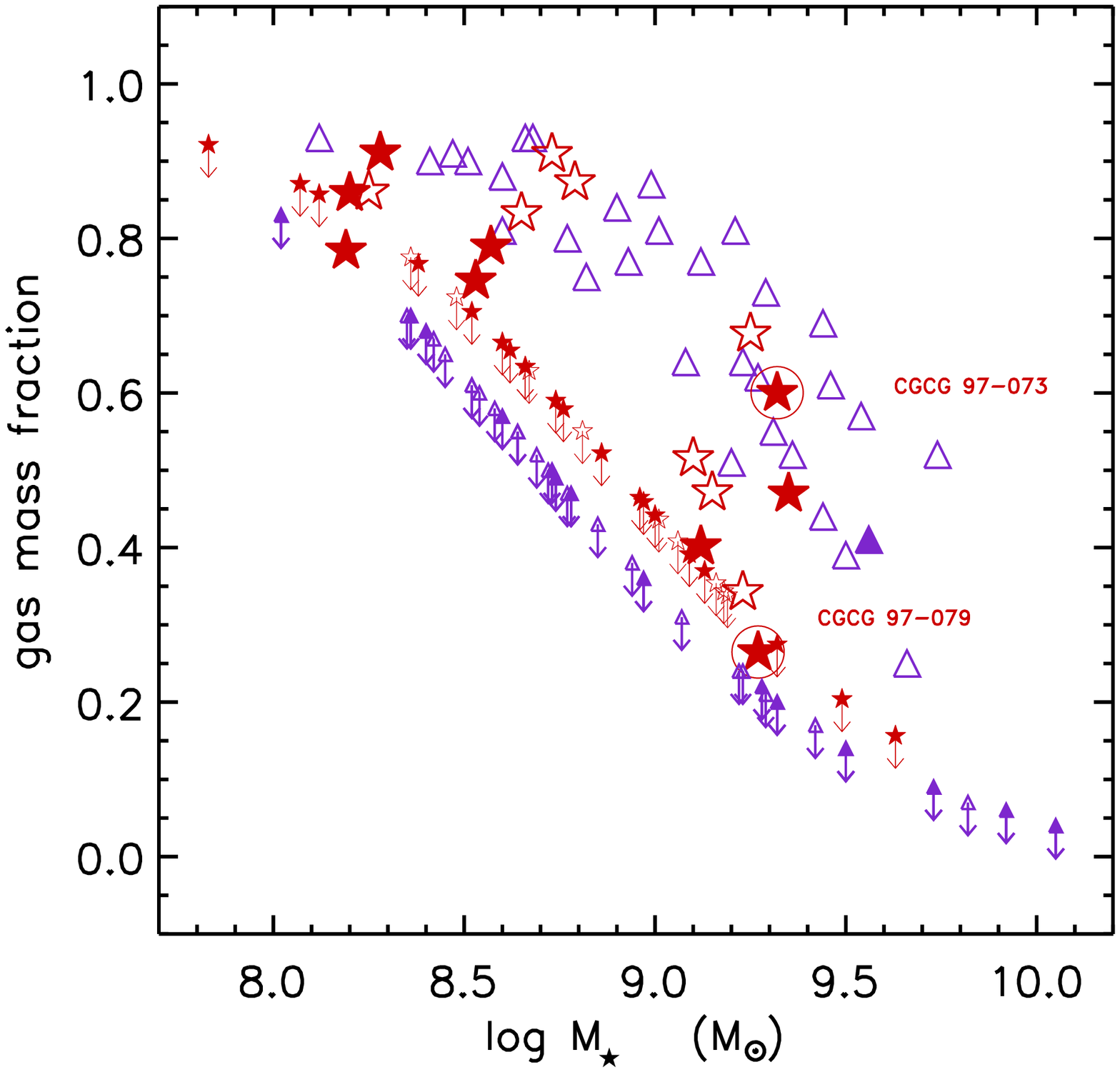}
\caption{Left: the oxygen abundance versus the gas mass fraction for  A1656 (triangles) and A1367 (stars, filled symbols mean $R\le R_{200}$), for the galaxies within the covered regions by AGES and ALFALFA. The green continuous line indicates the theoretical yield $y_o=0.0074$, and the blue dashed line corresponds to a lower yield $y_o=0.002$. Right: the gas mass fraction $\mu$ versus the stelar mass. Small points in both plots represent  HI mass upper limits for the galaxies included in the surveyed regions but not having HI measurements. The arrows indicate the dirrection to which these upper limits could be displaced. \label{fig12}}
\end{figure*}

%% file: fig13.tex
\begin{figure*}
\center
\includegraphics[width=12cm]{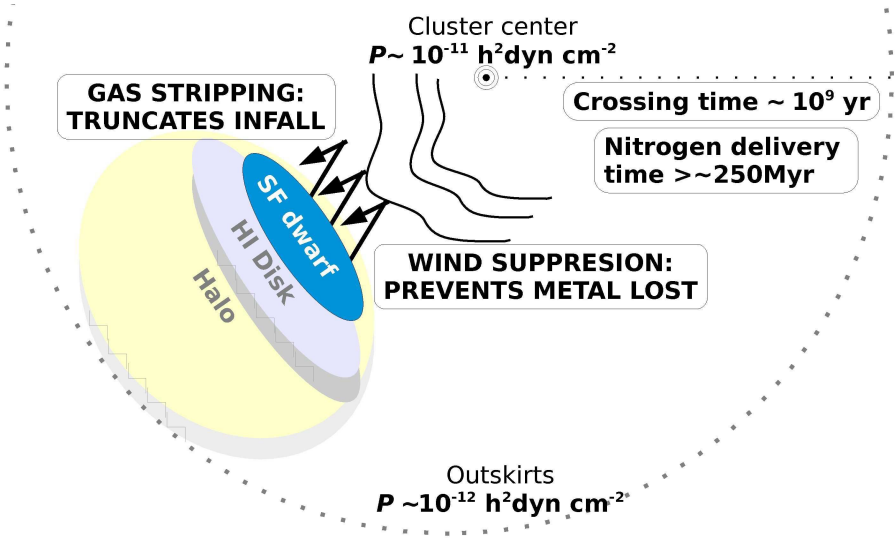}
\caption{A schematic view of the main mechanisms that could affect the chemical evolution of a low-mass galaxy in a massive cluster like Coma. As this galaxy enters the cluster environment, the ISM-ICM interaction produces star-burst events, accelarating gas depletion as compared to its isolated counterparts. The pressure of the ICM (depicted as pleats), in the region inside $R_{200}$ is high enough to produce wind suppresion, and the reaccretion of the wind material (the returning arrows) prevents dwarf galaxies from loosing the produced metals. Additionally, the accretion rate of pristine gas becomes truncated, as the HI disk and the hot halo reservoir of the dwarf galaxies are effectively stripped in that region.\label{fig13}}
\end{figure*}